\documentclass{aastex63}

\usepackage{graphicx,amssymb,amstext,amsmath}
\usepackage{latexsym}
\usepackage{multirow}
\usepackage{scalerel}
\usepackage[makeroom]{cancel}
\usepackage{color, colortbl}
\usepackage{enumitem}

\DeclareMathOperator{\sgn}{sgn}

\definecolor{Gray}{gray}{0.95}

\received{23 March 2021}
\revised{9 May 2021}
\accepted{21 May 2021}

\submitjournal{AJ}

\shorttitle{Multicolor Variability of Young Stars in the Lagoon Nebula}
\shortauthors{Venuti, L. et al.}

\begin{document}

\title{Multicolor Variability of Young Stars in the Lagoon Nebula: Driving Causes and Intrinsic Timescales}

\correspondingauthor{Laura Venuti}
\email{laura.venuti@nasa.gov}

\author{Laura Venuti}
\altaffiliation{NASA Postdoctoral Program Fellow}
\affiliation{NASA Ames Research Center, Moffett Blvd., Mountain View, CA 94035, USA}

\author{Ann Marie Cody}
\affiliation{SETI Institute, 189 Bernardo Ave, Suite 200, Mountain View, CA 94043, USA}


\author{Luisa M. Rebull}
\affiliation{Infrared Science Archive (IRSA), IPAC, 1200 E. California Blvd., California Institute of Technology, Pasadena, CA 91125, USA}

\author{Giacomo Beccari}
\affiliation{European Southern Observatory, Karl-Schwarzschild-Strasse 2, 85748 Garching bei M\"unchen, Germany}

\author{Mike Irwin}
\affiliation{Institute of Astronomy, University of Cambridge, Madingley Road, Cambridge, CB3 0HA, UK}

\author{Sowmya Thanvantri}
\affiliation{University of California, Berkeley, 101 Sproul Hall, Berkeley, CA 94720, USA}

\author{Steve B. Howell}
\affiliation{NASA Ames Research Center, Moffett Blvd., Mountain View, CA 94035, USA}

\author{Geert Barentsen}
\affiliation{Bay Area Environmental Research Institute, 625 2nd St., Suite 209, Petaluma, CA 94952, USA}



\begin{abstract}
Space observatories have provided unprecedented depictions of the many variability behaviors typical of low-mass, young stars. However, those studies have so far largely omitted more massive objects ($\sim$2\,$M_\odot$ to 4--5\,$M_\odot$), and were limited by the absence of simultaneous, multi-wavelength information. We present a new study of young star variability in the $\sim$1--2~Myr-old, massive Lagoon Nebula region. Our sample encompasses 278 young, late-B to K-type stars, monitored with \textit{Kepler/K2}. Auxiliary $u,g,r,i,H\alpha$ time series photometry, simultaneous with \textit{K2}, was acquired at the Paranal Observatory. We employed this comprehensive dataset and archival infrared photometry to determine individual stellar parameters, assess the presence of circumstellar disks, and tie the variability behaviors to inner disk dynamics. We found significant mass-dependent trends in variability properties, with B/A stars displaying substantially reduced levels of variability compared to G/K stars for any light curve morphology. These properties suggest different magnetic field structures at the surface of early-type and later-type stars. We also detected a dearth of some disk-driven variability behaviors, particularly dippers, among stars earlier than G. This indicates that their higher surface temperatures and more chaotic magnetic fields prevent the formation and survival of inner disk dust structures co-rotating with the star. Finally, we examined the characteristic variability timescales within each light curve, and determined that the day-to-week timescales are predominant over the \textit{K2} time series. These reflect distinct processes and locations in the inner disk environment, from intense accretion triggered by instabilities in the innermost disk regions, to variable accretion efficiency in the outer magnetosphere.
\end{abstract}




\section{Introduction} \label{sec:intro}

The few million years (Myr) age mark represents a crucial juncture in stellar evolution, when newly formed protostars emerge from their nebular cocoon, become optically visible, and enter the pre-main sequence (PMS) track. The physics of young stellar objects (YSOs) during these early stages is dominated by the interaction between the central star and the surrounding disk of gas and dust, a ubiquitous outcome of the star formation process \citep[e.g.,][]{shu1987, mckee2007}. The exchange of mass and angular momentum between the star and the disk, regulated via the process of magnetospheric accretion \citep[e.g.,][]{hartmann2016}, has a profound impact on the long-term evolution of the star. The dynamics of star-disk interaction is also directly relevant to the process of planet formation, by triggering changes in the local disk structure \citep[e.g.,][and references therein]{morbidelli2016}, and by influencing planetary migration and the location of ``planet traps'' across the disk \citep{romanova2019}.

While the last few years have witnessed a true revolution in protoplanetary disk surveys with the Atacama Large Millimeter/submillimeter Array (ALMA; e.g., \citealp{ansdell2016, barenfeld2016}), the spatial scales of the inner disk region relevant to magnetopsheric star-disk interactions ($\lesssim$0.1~AU; \citealp{dullemond2010}) are hard to resolve with current facilities, even for the YSOs closest to us \citep{andrews2016}. To date, mapping the star-disk emission across the spectrum, and tracing the variability of the observed emission features, represent the most direct probes to disclose the physics of the star-inner disk environment. In particular, {observations in the ultraviolet (UV) reveal any energetic emission from accretion shocks that form when streams of material are channeled from the inner disk onto the star \citep[e.g.,][and references therein]{calvet1998,gullbring1998,schneider2020}. The accelerated gas in magnetospheric accretion funnels are also associated with distinctive spectroscopic signatures such as H$\alpha$ line emission \citep[e.g.,][]{white2003,kurosawa2006}. Observations in the optical, sensitive to the photospheric emission from the central star, allow us to reconstruct any modulation effects by surface features \citep[e.g.,][]{bouvier1995,grankin2008} or circumstellar structures \citep[e.g.,][]{bouvier2007,fonseca2014,frasca2020}. Finally, infrared (IR) wavelengths trace the thermal emission produced by dust at different locations within the disk, from the inner AU (near-IR; e.g., \citealp{robberto2010,roquette2020}), to radii of few AU (mid-IR; e.g., \citealp{oliveira2004,gunther2014}) to tens of AU (far-IR; e.g., \citealp{fedele2013,alonso_martinez2017}).}

Although young stars have long been known to be strongly variable sources \citep[e.g.,][]{joy1945}, {and although numerous photometric and spectroscopic monitoring campaigns of young stars have been conducted from ground-based facilities down to timescales of minutes \citep[e.g.,][]{bastian1979a,bastian1979b},} time series data before the era of space observatories did not possess the cadence, duration, and precision required to categorize different YSO behaviors beyond the assertion of periodic vs. irregular vs. extinction-driven variability patterns \citep{herbst1994}. The paradigm began to shift with the Microvariability and Oscillations of Stars (MOST) telescope \citep{walker2003}, the {\it Spitzer} Space Telescope \citep{werner2004}, and the Convection, Rotation, and planetary Transits ({\it CoRoT}) telescope \citep{baglin2003, auvergne2009}. The accuracy and homogeneity of space data enabled showcasing YSO variability as a panchromatic phenomenon \citep[e.g.,][]{morales_calderon2011}, that can appear with characteristic features on timescales as short as hours \citep[e.g.,][]{rucinski2008}. {Although not sensitive to the eruptive variability events typical of EXors and FUors, which take place over years-long timescales \citep{audard2014}, those space-based, high-cadence datasets enabled detailing, for the first time, the distinct hour-to-month} light curve morphology classes {characteristic of young, disk-bearing stars} \citep{alencar2010}, {and brought to light} a larger variety of behaviors than could be appreciated from the ground.\looseness=-1

The current framework for time domain studies of young stars was first established by the Coordinated Synoptic Investigation of NGC~2264 (CSI~2264; \citealp{cody2014}). The campaign employed {\it CoRoT} and {\it Spitzer}, in coordination with a dozen other space- and ground-based observatories, to monitor the hours-to-weeks variability of hundreds of low-mass YSOs in the 3--5~Myr-old cluster NGC~2264 \citep{dahm2008}. The quality of space-based data enabled the implementation of quantitative metrics to classify light curve behaviors and their occurrence rates. Multiwavelength data gathered from the ground in the UV \citep{venuti2014, venuti2015} and in the H$\alpha$ band \citep{sousa2016} enabled assessing the connection between distinct variability features and disk-related phenomena. The original census of YSO behaviors was expanded from a few categories to at least eight, including two separate irregular patterns, named bursters \citep[dominated by brightening events;][]{stauffer2014} and stochastic \citep{stauffer2016}, and a variety of morphologies for light curves dominated by fading events, the so-called dippers \citep{stauffer2015, mcginnis2015}. Bursting and stochastic behaviors were interpreted as the imprint of intense, unstable, and/or time variable accretion, as earlier predicted from a theoretical standpoint by \citet{kulkarni2008} and \citet{kurosawa2013}. Dipping behaviors were interpreted as the result of repeated partial occultations of the stellar surface by intervening inner disk warps, or by dust entrained in accretion columns, in star-disk systems observed at mid-to-high inclinations (i.e., close to edge-on). Regular, \mbox{(quasi-)}periodic behaviors, also common among the investigated YSOs, were ascribed to modulation by dark magnetic spots or bright accretion spots at the stellar surface. 

The paradigm for YSO variability that emerged from CSI~2264 was later tested and expanded in studies of (among others) the $\sim$8~Myr-old Upper Scorpius \citep{preibisch2008} and the $\sim$2~Myr-old $\rho$ Ophiucus \citep{wilking2008}, targeted as part of the {\it K2} mission \citep{howell2014} with the {\it Kepler} spacecraft \citep{borucki2010}. Such studies led to: i) identifying new subclasses among periodic variables, possibly driven by warm gas clouds co-rotating with the star after the inner disk has been cleared \citep{stauffer2017}; ii) establishing a homogeneous pattern of rotational evolution across the PMS \citep{rebull2018,rebull2020}; iii) characterizing the duty cycle, duration, and timescales of bursting events in irregular variables, in connection with different models of episodic accretion \citep{cody2017}; iv) exploring YSO variability properties in relation to inner disk radii and outer disk inclinations \citep{cody2018}, in synergy with ALMA surveys of protoplanetary disks \citep{carpenter2014, barenfeld2017}. However, those results were obtained mainly for low-mass YSOs ($\lesssim 1.5\, M_\odot$; spectral type K and M). Higher mass young stars are known to be variable as well \citep[e.g.,][]{teixeira2018}, but the frequencies and detailed morphologies of their light curve types have yet to be explored. Moreover, our understanding of YSO behaviors, as shaped during CSI~2264 and early {\it K2} campaigns, pivots on the role of the stellar magnetic field, both in driving spot-modulated variability and in governing accretion-dominated variability. However, while magnetic fields are detected ubiquitously among low-mass YSOs, they are rarely found on intermediate-mass YSOs and Herbig Ae/Be stars that lack a convective envelope \citep{villebrun2019}. This may suggest that different variability properties or mechanisms are to be expected for higher mass YSOs. Another limitation of earlier {\it K2} surveys was the absence of auxiliary multiwavelength data, gathered simultaneously. Indeed, while space data provide an unprecedented view of the morphology of flux variations exhibited by YSOs on various timescales, multiwavelength information is crucial to reconstruct the spectrum of the luminosity variations and to pinpoint their physical origin (e.g., \citealp[][and references therein]{vrba1993, venuti2015}).\looseness=-1

In this work, we address the issues listed above by conducting a thorough analysis of YSO variability in the Lagoon Nebula region, a rich H{\scriptsize II} region located in the Sagittarius Arm of our Galaxy \citep{tothill2008}. Much of its PMS population is comprised by the open cluster NGC~6530, an extremely young ($0.7^{+0.9}_{-0.4}$~Myr; \citealp{prisinzano2019}) star-forming region situated in the heart of the Lagoon Nebula, at a distance of $\sim$1325~pc \citep{damiani2019}. The region hosts a comparatively higher mass population than other young clusters studied earlier, including numerous O and B stars. The Lagoon Nebula was observed with {\it Kepler} as part of the {\it K2} Campaign~9, and other telescopes were employed at the same time to sample the variability of Lagoon Nebula YSOs at different wavelengths: the Very Large Telescope (VLT) of the European Southern Observatory (ESO), with its spectrograph FLAMES (Fibre Large Array Multi Element Spectrograph; \citealp{pasquini2000}), to probe the spectroscopic H$\alpha$ variability; and the VLT Survey Telescope (VST), with its wide-field imager OmegaCAM \citep{kuijken2011}, to probe the color variability of Lagoon Nebula YSOs in $u,g,r,i$ and narrow-band $H\alpha$ filters. We use here the {\it K2} and VST/OmegaCAM data to measure the amount of multicolor variability and to reconstruct the color behaviors exhibited by different categories of YSO variables in our sample, in relation to stellar and circumstellar properties. The analysis of spectroscopic variability is deferred to a later work.

The paper is organized as follows. Section~\ref{sec:targets_data} describes our target selection, the time series photometric data collected with {\it Kepler/K2} and VST/OmegaCAM, and the literature information used to complement our own dataset (notably in the IR). In Sect.\,\ref{sec:stellar_parameters_classification}, we illustrate the methods used to derive individual stellar properties (extinction, spectral types) and to identify disk-bearing and disk-free YSOs in our sample. In Sect.\,\ref{sec:YSO_variability}, we examine the variability properties exhibited by Lagoon Nebula YSOs in {\it K2} and VST/OmegaCAM filters: the morphological classification of {\it K2} light curves as a function of disk status and spectral type (Sect.\,\ref{sec:K2_lc_class}); the correlated variability exhibited by young stars in our sample at different wavelengths ($u,r,i,H\alpha$), as a function of luminosity (Sect.\,\ref{sec:VST_variability}); and the connection between {\it K2} variability types and stellar colors at UV, optical, and IR wavelengths, which trace disk-related phenomena (Sect.\,\ref{sec:var_class_colors}). In Sect.\,\ref{sec:correlated_lumin_color_var}, we investigate the color variation trends associated with prominent flux variability features in the {\it K2} light curves for disk-bearing YSOs, and discuss their implications for the magnetospheric accretion and star--disk interaction structure around different types of variables. In Sect.\,\ref{sec:SF}, we conduct a structure function analysis of the {\it K2} light curves to extract the characteristic timescales of variability for all YSO categories in our sample, and we explore the dominant types of signal (white noise, flicker noise, Brownian noise, sinusoidal variations) that govern the time series, and the time spans on which they emerge. In Sect.\,\ref{sec:discussion} we discuss our results in relation to mass-dependent stellar internal structure, circumstellar dust structure, and magnetic field properties; we also examine their implications for the stability of the circumstellar environment at, and beyond, the inner disk. Our conclusions are summarized in Sect.\,\ref{sec:conclusions}.

\section{Target selection and observational data} \label{sec:targets_data}

\subsection{The Lagoon Nebula region in the literature} \label{sec:membership}

The Lagoon Nebula is a benchmark for investigations of early stellar evolution and to probe the conditions for planet formation, thanks to the very young age of the NGC~6530 cluster, and to its rich population of PMS stars that span a wide range of masses. As such, the region has been extensively studied over the past decades with a variety of diagnostics, and over 35 published papers or survey catalogs exist to date that encompass the NGC~6530 region. These include: the Massive Young star-forming complex Study in Infrared and X-ray (MYStIX; \citealp{kuhn2013}); the Gaia survey \citep{damiani2019}, the Gaia-ESO Survey (GES; \citealp{prisinzano2019,wright2019}), the Panoramic Survey Telescope and Rapid Response System (Pan-STARRS; \citealp{chambers2016,flewelling2016}), and the VST Photometric H$\alpha$ Survey of the Southern Galactic Plane and Bulge (VPHAS+; \citealp{kalari2015}) in the optical; and the UKIRT Infrared Deep Sky Survey (UKIDSS; \citealp{lawrence2007}), the Two Micron All Sky Survey (2MASS; \citealp{skrutskie2006}), {\it Spitzer} data including the Galactic Legacy Infrared Mid-Plane Survey Extraordinaire (GLIMPSE; \citealp{churchwell2009}) and the Spitzer Enhanced Imaging Products (SEIP) archive, and Wide-field Infrared Survey Explorer (WISE; \citealp{wright2010}) data in the IR. We cross--matched all of the existing Lagoon Nebula catalogs in the literature, to build an extensive sample of probable and candidate members based on any diagnostics of stellar youth or PMS status (which include enhanced X-ray emission related to magnetic activity, UV and H$\alpha$ emission related to accretion processes, IR flux emission related to thermal emission from the disk, lithium absorption indicative of youth, and astrometric association with the cluster). In total, we found {$\sim$3000 YSO candidates, out of which} $\sim$1000--1500 candidate members projected onto the area covered by the {\it K2} mosaic. This number includes very faint or embedded sources, detected only in the IR. Around 700 candidate members possess an optical counterpart in the literature, and have corresponding {\it Kepler} magnitudes ranging from 8 to 20 (spectral types B to M, or masses between 5 and 0.2~$M_\odot$). After filtering down in magnitude to ensure sufficient signal-to-noise ratio in {\it K2} photometry, we retained a list of $\sim$300 YSOs, primarily located toward the NGC~6530 cluster core.

\subsection{{\it K2} observations} \label{sec:K2}

\begin{figure}
\centering
\includegraphics[width=0.6\textwidth]{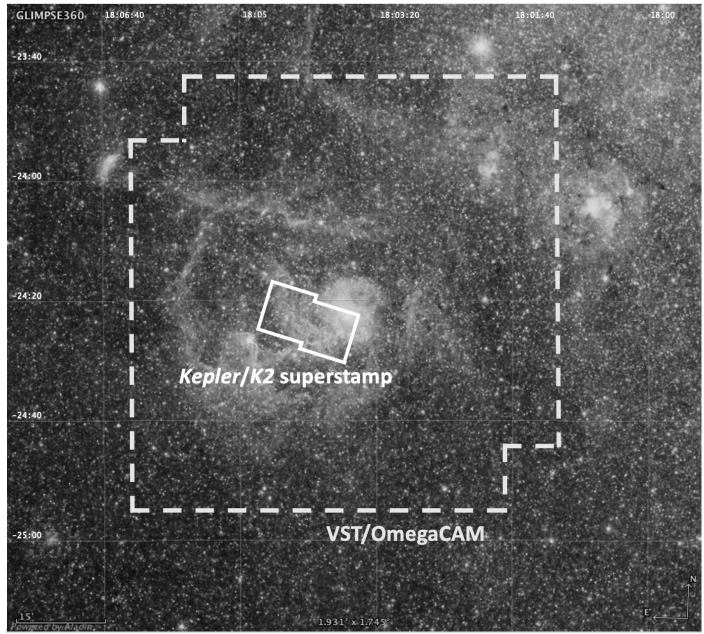}
\caption{Black and white image of the Lagoon Nebula field from {\it Spitzer}/GLIMPSE, over which the total area covered with VST/OmegaCAM (sum of two dithered pointings) and the {\it Kepler}/{\it K2} superstamp region are outlined as dashed and solid contours, respectively. The original version of the background image was prepared with Aladin Desktop \citep{bonnarel2000}.}
\label{fig:Lagoon_FoVs}
\end{figure}

{\it K2} monitored the Lagoon Nebula from April 21 through July 1, 2016. The campaign was split into two segments, with an interruption of nearly four days between May 18 and 22. The bulk of the NGC~6530 cluster was observed as a {\em K2} ``superstamp" region (see Fig.\,\ref{fig:Lagoon_FoVs}), for which a $\sim$15\arcmin$\times$9\arcmin\ image was acquired every 6.5 seconds. Sets of 270 frames were then co-added for an ultimate cadence of roughly 30 minutes. Additional data was acquired simultaneously for 11 stars outside the superstamp region. 

To generate light curves, we performed photometry with circular apertures ranging in radii from one to four pixels. {\em Kepler's} pointing was unstable on $\sim$6-hour timescales during the {\em K2} mission, and thus it is important to re-center the apertures in each successive image. For superstamp targets (i.e., the majority of our targets), we accomplished this by converting stars' known right ascension and declination into pixel position using an accurate world coordinate system (WCS) transformation, as described by \citet{cody2018rnaas}. For non-superstamp targets, we
computed a flux-weighted centroid as described in \citet{cody2018}.

Once an aperture was centered, we summed the flux and subtracted out the median background as determined in a 10$\times$10~pixel region surrounding the source center, with outliers iteratively removed. After performing the aperture photometry, we selected the preferred light curve by visually examining the set of four produced for each star. If there were no nearby companions on the image, then we chose the aperture that produced the least noisy light curve. However, if there were one or more close companions, we
chose the aperture least subject to flux contamination. In cases for which contamination could not be avoided and the origin of variability was unknown, we eliminated that object from our sample. This process left us with a total of 278 stars with sufficient quality {\em K2} light curves. This sample of Lagoon Nebula YSOs, which is nearly complete down to magnitudes V$\sim$16.5--17, constitutes the focus of our work and is  listed in Table~\ref{tab:Lagoon_YSOs_list}.

\subsection{VST/OmegaCAM dataset} \label{sec:VST}

The VST/OmegaCAM run on the Lagoon Nebula field was executed over a period of 3.5 weeks (June 16 through July 10, 2016; Program ID 297.C-5033(A), PI Cody), partly overlapping with the second half of the {\it K2} run. Seventeen observing epochs were obtained, distributed over 14 non-consecutive days. Each observing block comprised two dithered exposures in the Sloan Digital Sky Survey (SDSS) {\it u,g,r,i} filters (Fig.\,\ref{fig:Lagoon_FoVs}) and three dithered exposures in a narrow-band, 659-nm H$\alpha$ filter \citep{drew2014}. Each set of exposures per filter was repeated twice, once with longer exposure times (70\,s in $u$, 5\,s in $g$, 4\,s in $r$ and $i$, and 30\,s in H$\alpha$), and once with very short exposure times {(1\,s in $u$, 0.3\,s in $g$, $r$ and $i$, and 0.5\,s in H$\alpha$), to recover the brightest (B-type) stars in our field that would be saturated in exposures of a few seconds or less, particularly in the red filters}. On the two nights when more than one observing epoch were acquired, the distinct epochs were separated by a lag of 20 minutes to 1~hour. All frames were processed using the Cambridge Astronomy Survey Unit (CASU) pipeline, and point source catalogs were produced for each band and exposure. All single-epoch catalogs were then cross-matched to produce light curves. Only sources that were detected in at least ten separate exposures were retained for the light curve production. A total of 187\,530 sources were retrieved, with light curves in at least one VST band; among these, 2444 are included in the list of candidate YSOs compiled as described in Sect.\,\ref{sec:membership}.

The instrumental photometry was calibrated to SDSS magnitudes in the Vega system by cross-correlating the list of average magnitudes for all point sources in our VST survey with the catalog of the Lagoon Nebula region published by the consortium of the VPHAS+ project, which was conducted using the same instrument. The photometric calibration was ultimately performed on a sample of 29\,492 objects, common to our VST catalog and to the VPHAS+ catalog, and comprising only field stars (not YSO candidates). No significant color effects were observed, hence the calibration was performed band by band to correct only for zero-point offsets. For each filter, a 10\,$\sigma$-clipping routine was performed to reject outliers affected by very large photometric discrepancies between our catalog and the VPHAS+ catalog. Then, the zero-point correction was calculated as the typical magnitude difference measured, across the remaining sample, between our catalog and the VPHAS+ catalog, while the root-mean square (rms) dispersion around this value was extracted as uncertainty on our zero-point correction. The resulting magnitude corrections are listed {in Eq.\,\ref{eqn:VST_phot_calib}, and the average calibrated photometry for the sample of YSOs investigated in this study is reported in Table~\ref{tab:Lagoon_YSOs_list}}.

\begin{equation} \label{eqn:VST_phot_calib}
\begin{aligned}
u - u_{\scaleto{VPHAS+}{4pt}} = 7.55\pm0.07 \\
g - g_{\scaleto{VPHAS+}{4pt}} = 5.30\pm0.04 \\
r - r_{\scaleto{VPHAS+}{4pt}} = 5.43\pm0.03 \\
i - i_{\scaleto{VPHAS+}{4pt}} = 5.88\pm0.05 \\
H\alpha - H\alpha_{\scaleto{VPHAS+}{4pt}} = 8.31\pm0.09
\end{aligned}
\end{equation}

\begin{figure}
\centering
\includegraphics[width=\textwidth]{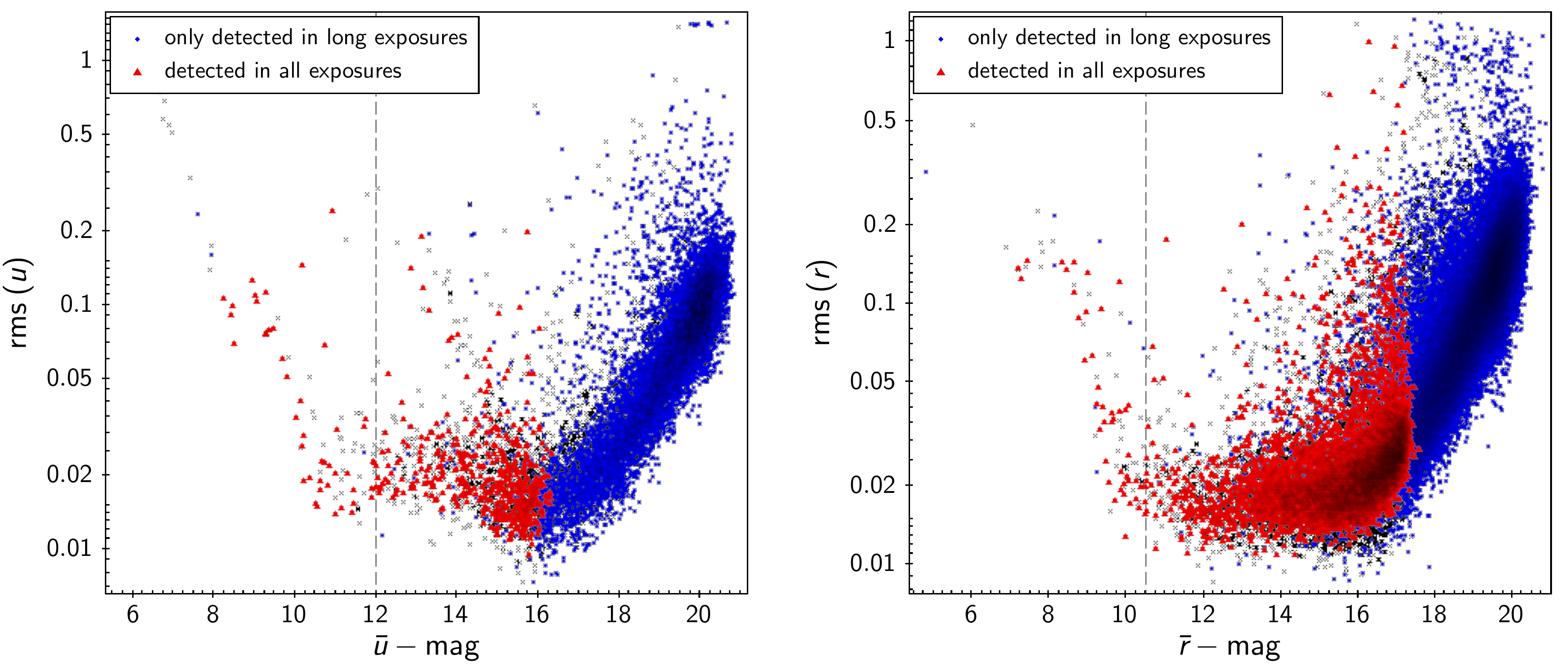}
\caption{Rms dispersion measured in VST/OmegaCAM $u$-band (left) and $r$-band (right) light curves for all point sources in the Lagoon Nebula field, as a function of the average light curve magnitude. Objects marked as blue diamonds correspond to sources that were detected only in the long exposures (see Sect.\,\ref{sec:VST}), while objects marked as red triangles correspond to sources detected in all exposures of each observing block. The dashed gray line marks the estimated saturation limit below which sources were discarded.}
\label{fig:VST_ur_rms}
\end{figure}

{
Since the processes we aim to investigate (dynamics of disk accretion, star--disk interaction, stellar activity) develop primarily over timescales from hours to weeks \citep[e.g.,][]{grankin2008, costigan2014}, we decided to retain, in our VST time series analysis, only the error-weighed average magnitudes measured for each observing epoch in each VST filter, in order to reduce the impact of photometric scatter within individual observing blocks. The final VST time series used for the analysis therefore comprise 17 photometric measurements, taken at separations between $\sim$0.5~hours and five days from one another. In addition, plots of the light curve rms vs. average magnitude were inspected in all filters to identify the brightness limit above which the trend starts to diverge from the expected behavior of decreasing light curve noise for increasing brightness \citep[e.g.,][]{moraux2013,venuti2015}. Examples of such plots for the $u$-band and the $r$-band are shown in Fig.\,\ref{fig:VST_ur_rms} (left panel and right panel, respectively). 
}

{
For magnitudes $\overline{u} \gtrsim 16$, the measured light curve rms scatter increases exponentially for increasing (fainter) magnitudes, consistent with expectations. A trend reversal is observed for $\overline{u} < 16$, where the measured rms slightly increases for decreasing magnitudes and then remains approximately constant down to $\overline{u} \sim 12$. This trend reversal can be explained by the fact that objects fainter than $\overline{u} \sim 16$ were only detected in the long exposures, while objects brighter than $\overline{u} \sim 16$ were mostly detected in all exposures, with potentially different accuracy depending on the exposure time. A similar effect, albeit less pronounced, can be observed on the rms\,($r$) vs. $\overline{r}$ diagram. Nevertheless, these distinct behaviors between brighter stars and fainter stars do not affect our statistical analysis of variability described later, as we have only considered error-weighed average magnitudes for each exposure block (composed of two dithered long exposures and two dithered short exposures), as indicated above. Moreover, in our statistical assessment of variability properties (see, in particular, Sect.\,\ref{sec:VST_variability}), each group of objects (e.g., disk-bearing YSOs, disk-free YSOs) was weighted against other stellar groups (field stars, differents types of YSOs) in the same magnitude range, therefore ensuring a self-consistent analysis on a relative scale. On the other hand, Fig.\,\ref{fig:VST_ur_rms} also shows a tail of bright objects, at $\overline{u} \leq 12$ and $\overline{r} \leq 10.5$, which appear to diverge from the trend traced by stars down to magnitudes $\sim$15. We suspect that these objects may be close to, or past the saturation limit, and therefore discarded all sources to the left of the dashed gray lines on the diagrams. Similar cuts were applied in the other bands, at $\overline{g} = 11$, $\overline{i} = 11$, and $\overline{H\alpha} = 10.5$.
}

\subsection{Additional catalogs used for this work}

In order to reconstruct the properties of the circumstellar environment for our target stars, we complemented our optical/UV datasets with IR data from literature surveys mentioned in Sect.\,\ref{sec:membership}. In the near-IR, we gathered $J,H,K$ data from the 2MASS and UKIDSS catalogs, which encompass $\sim$90\% of the Lagoon Nebula YSOs for which {\it K2} photometry could be extracted. In the mid-IR, we gathered data obtained with the Infrared Array Camera (IRAC) onboard {\it Spitzer} at 3.6\,$\mu$m, 4.5\,$\mu$m, 5.8\,$\mu$m, and 8.0\,$\mu$m, available for 75\% of the {\it K2} sample \citep[][and references therein]{kumar2010, povich2013, broos2013}. More specifically, to assign IRAC magnitudes to each of our targets, we prioritized extracting all four photometric measurements in IRAC bands from the same source for each given star, and used, in the order, the data released by SEIP, MYStIX, GLIMPSE, or \citet{kumar2010}. When no IRAC photometry from any sources was available, we downloaded the SEIP images of our targets' field of view and extracted our own aperture photometry measurements (Rebull et al., in prep.). As a consistency check, we extended this procedure to targets with photometry already provided in the archival catalogs, and we could ascertain that the derived magnitudes typically agree within 0.05--0.10 mag. We also gathered available data from the WISE catalog at wavelengths from 3.4\,$\mu$m to 22\,$\mu$m, albeit only for a quarter of our target list. We used this IR dataset to sort our targets into disk-bearing and disk-free sources, as detailed in Sect.\,\ref{sec:disk_class}. In addition, we collected optical photometry in standard Johnson-Cousins filters (notably V, R, and I; e.g., \citealp{vandenancker1997, sung2000, prisinzano2005}), which we used for the selection of IR excess sources, as also described in Sect.\,\ref{sec:disk_class}.

\section{Stellar and circumstellar properties of young stars in the Lagoon Nebula} \label{sec:stellar_parameters_classification}

Of the 278 objects in our {\em K2} sample, each has a counterpart in the VST catalog, with a median separation of 0.15$''$ and a maximum separation of 1.27$''$ from the spatial coordinates of the {\it K2} detection. We used the collected multi-wavelength photometry to inspect the location of each object on various color-color diagrams, in order to evaluate key stellar properties such as spectral type and to assess the disk status of each YSO, as reported in the following. 

\begin{deluxetable*}{c c c c c c c c c c c}
\centerwidetable
\tablecaption{Young stars targeted during {\it K2} Campaign 9 in the Lagoon Nebula core, and multi-band $u,g,r,i,H\alpha$ photometry in the Vega system gathered for each source with VST/OmegaCAM.}
\label{tab:Lagoon_YSOs_list}
\tablehead{\colhead{EPIC ID\tablenotemark{a}} & \colhead{2MASS ID} & \colhead{Spectral class\tablenotemark{b}} & \colhead{$A_V$} & \colhead{Disk flag\tablenotemark{c}} & \colhead{$u$-mag} & \colhead{$g$-mag} & \colhead{$r$-mag} & \colhead{$i$-mag} & \colhead{$H\alpha$-mag} & \colhead{{\it K2} class\tablenotemark{d}}}
\startdata
 224366753 & J18034826-2422233 & Late-G & 0.5 & ? & 17.04 & 15.52 & 14.49 & 14.01 & 14.09 & P \\
 224355341 & J18034935-2423374 & Mid-K & 0.3 & N & 19.22 & 17.47 & 16.18 & 15.62 & 15.51 & P \\
 224378426 & J18035006-2421079 & Early-K & 0.1 & N & 17.28 & 15.22 & 14.17 & 13.72 & 13.99 & QPS \\
 224379624 & J18035063-2421001 & Late-B & 0.3 & N & 13.14 & 12.36 & 12.32 & 12.29 & 12.28 & U \\
 224361254 & J18035221-2422592 & Early-K & 0.5 & N & 16.25 & 14.34 & 13.23 & 12.71 & 12.98 & MP \\
\enddata
\tablenotetext{a}{Identification number for the target from the Ecliptic Plane Input Catalog, if available.}
\tablenotetext{b}{Approximate spectral class derived for each target as described in Sect.\,\ref{sec:Av_SpT}. The mention of ``early-" encompasses spectral subclasses 0 to 3, ``mid-" corresponds to subclasses 4 to 6, and ``late-" identifies subclasses 7 to 9.}
\tablenotetext{c}{Possible flag values are ``Y'' if the target is a disk-bearing YSO, ``N'' if the target is a disk-free object, and ``?'' if the spectral energy distribution of the object and the disk indicators described in Sect.\,\ref{sec:disk_class} provide contrasting or incomplete information. In our analysis, we consider objects labeled ``?'' as disk candidate YSOs.}
\tablenotetext{d}{Morphological class of the {\it K2} light curve for the target. Possible values for this field are ``P'' (periodic), ``QPS'' (quasi-periodic symmetric), ``S'' (stochastic), ``B'' (burster), ``QPD'' (quasi-periodic dipper), ``APD'' (aperiodic dipper), ``MP'' (multi-periodic), ``EB'' (eclipsing binary), ``N'' (non-variable), and ``U'' (unclassifiable), as detailed in Sect.\,\ref{sec:K2_lc_class}.}
\tablecomments{The Table is published in its entirety in the machine-readable format. A portion is shown here for guidance regarding its form and content.}
\end{deluxetable*}

\subsection{Individual extinction and spectral type estimates} \label{sec:Av_SpT}

\begin{figure*}
\centering
\includegraphics[width=\textwidth]{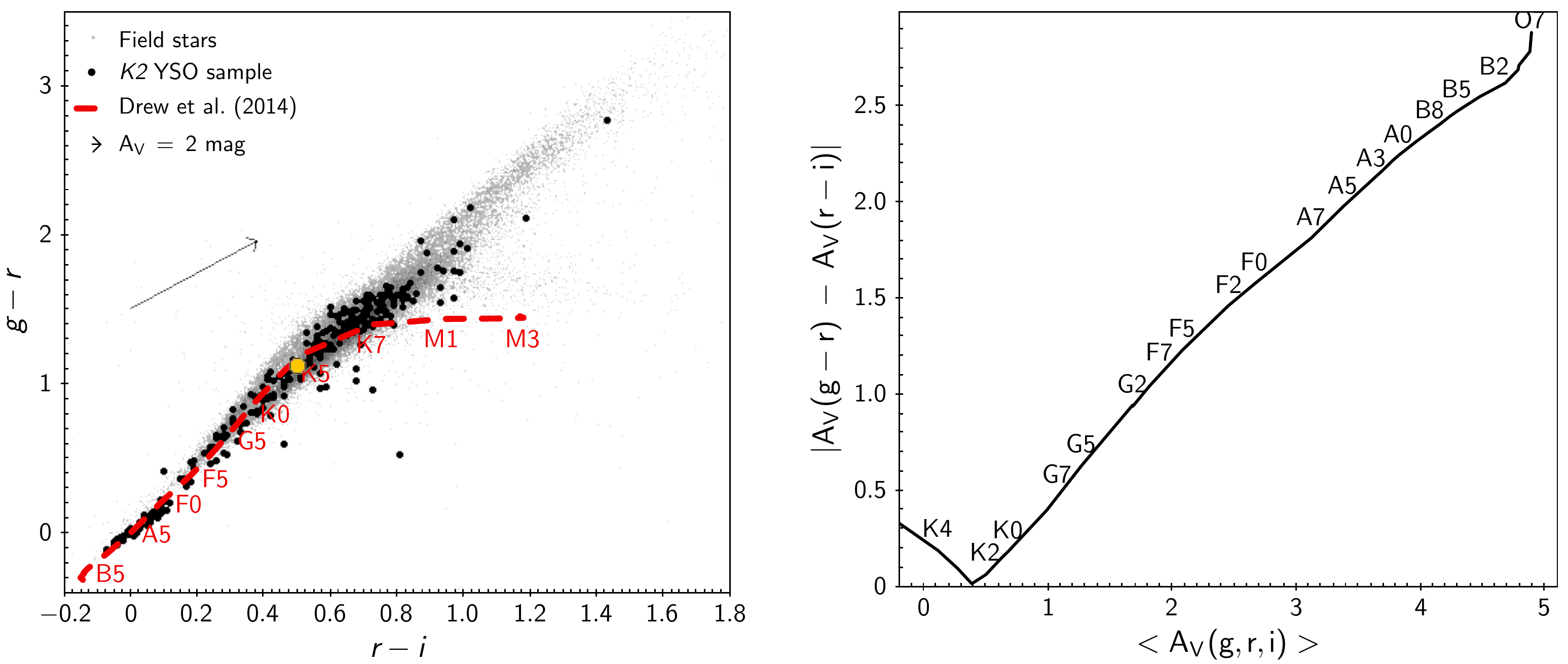}
\caption{{\it Left:} {Typical} VST $g,r,i$ colors of Lagoon Nebula members monitored with {\it K2} (black dots), overlaid on the color distribution of field stars (gray dots). The sequence of synthetic, non-reddened colors tabulated in \citet{drew2014} is dashed in red as a reference, and the location of representative SpTs is labelled along the sequence to guide the reader. The extinction vector shown on the diagram corresponds to $A_V = 2$~mag following a non-standard reddening law with $R_V = 5$. A yellow dot marks the object for which results of the $A_V$ derivation are shown as an example in the right panel. {\it Right}: Average $A_V$ measured from $r-i$ and $g-r$ colors, and absolute difference between the corresponding $A_V\,(r-i)$ and $A_V\,(g-r)$, calculated for the object marked in yellow on the left panel with respect to the non-reddened colors tabulated for each SpT in \citet{drew2014}, and labeled as a reference along the $A_V$ curve. The spectral type that minimizes the difference $|A_V\,(g-r) - A_V\,(r-i)|$ is extracted as best SpT estimate, and the corresponding average $<A_V\,(g,r,i)>$ is extracted as best estimate of individual $A_V$.}
\label{fig:ri_gr}
\end{figure*}

To assign estimates of individual extinction ($A_V$) and of spectral type (SpT) homogeneously across our sample, we used the $g,r,i$ colors provided by our VST dataset. We did not include $u$-band or $H\alpha$-band data for this step of the analysis, as they may be affected by accretion activity in these young sources. Similarly, we did not include IR data for the determination of stellar parameters, as they may be affected by thermal emission from the disk. To derive the best photometric estimate of A$_V$, and the corresponding photospheric SpT for each object, we assumed an anomalous reddening law $R_V = 5.0$ toward NGC~6530 members, as reported in \citet{prisinzano2019}, and adopted the sequence of synthetic, non-reddened colors for dwarfs tabulated in VPHAS+ bands by \citet{drew2014}, as shown in Fig.\,\ref{fig:ri_gr}.

For each object, we calculated what value of $A_V$ would be required to correct the observed $r-i$ and $g-r$ colors, in order to match the non-reddened colors tabulated for any given spectral subclass across the mid-O to mid-M range. This calculation was conducted independently on the two colors, using the non-standard reddening law mentioned earlier. We then selected as best SpT estimate the class for which the computed $A_V (r-i)$ and $A_V (g-r)$ exhibit the closest agreement. We estimated an rms uncertainty of 0.19~mag on the measured $A_V$, ensuing from the uncertainties on the photometry calibration (Eq.\,\ref{eqn:VST_phot_calib}). Therefore, negative $A_V$ solutions greater than -0.19 were considered consistent with $A_V\sim0$; conversely, solutions where the minimum absolute difference $|A_V\,(r-i) - A_V\,(g-r)|$ corresponds to an average $<A_V\,(g,r,i)>$ lower than -0.19, and those with minimum $|A_V\,(r-i) - A_V\,(g-r)|$ larger than 3$\times$rms = 0.57~mag, were discarded. 

{When implementing the procedure outlined above using the instantaneous $g-r$ and $r-i$ colors measured at the various VST monitoring epochs, individual estimates of the apparent spectral type can deviate by as much as one class above or below the typical range of SpT derived for each star. In order to assign a robust estimate of SpT to each object, we extracted a best-fit SpT from the average photometric properties detected for each source during the VST monitoring, and considered as uncertainty the typical dispersion around this value derived from individual observing epochs, mostly on the order of a few spectral subclasses.} By following this approach, we were able to assign a SpT estimate to $\sim$81\% of our sample: among these, $\sim$7\% were categorized as B-type stars, $\sim$13\% as A-type, $\sim$8\% as F-type, $\sim$17\% as G-type, and $\sim$55\% as K-type. {No SpT could be measured for the remaining sources due to a degeneracy between the observed color properties and the adopted extinction law, or because the best SpT estimates obtained in those cases from $r-i$ and $g-r$ colors did not agree within the accepted ranges discussed above.} 

Approximate spectral classes and estimated $A_V$ values for our targets, derived as described here, are reported in Table~\ref{tab:Lagoon_YSOs_list}.

\subsection{Selection of disk-bearing stars} \label{sec:disk_class}

To assess whether the young stars in our sample are surrounded by thick inner disks, or whether they have already cleared their close surroundings, we employed a variety of diagnostics. As a first step, we conducted a visual inspection of the spectral energy distribution (SED) of each object with a {\it K2} light curve, from UV to mid-IR wavelengths, to detect any excess emission at IR wavelengths above the flux level of the stellar photosphere \citep[e.g.,][]{robitaille2007}. This analysis resulted in a preliminary classification of our objects as Class~II sources (which exhibit significant excess emission at wavelengths longer than $\sim$2~$\mu$m, originated in a circumstellar disk) or Class~III sources (with little or no flux contribution from circumstellar dust), as originally proposed by \citet{lada1987}. We then corroborated our classification by investigating the IR color properties of our stars with respect to several disk indicators, as enumerated below.

\begin{figure}
\centering
\includegraphics[width=\textwidth]{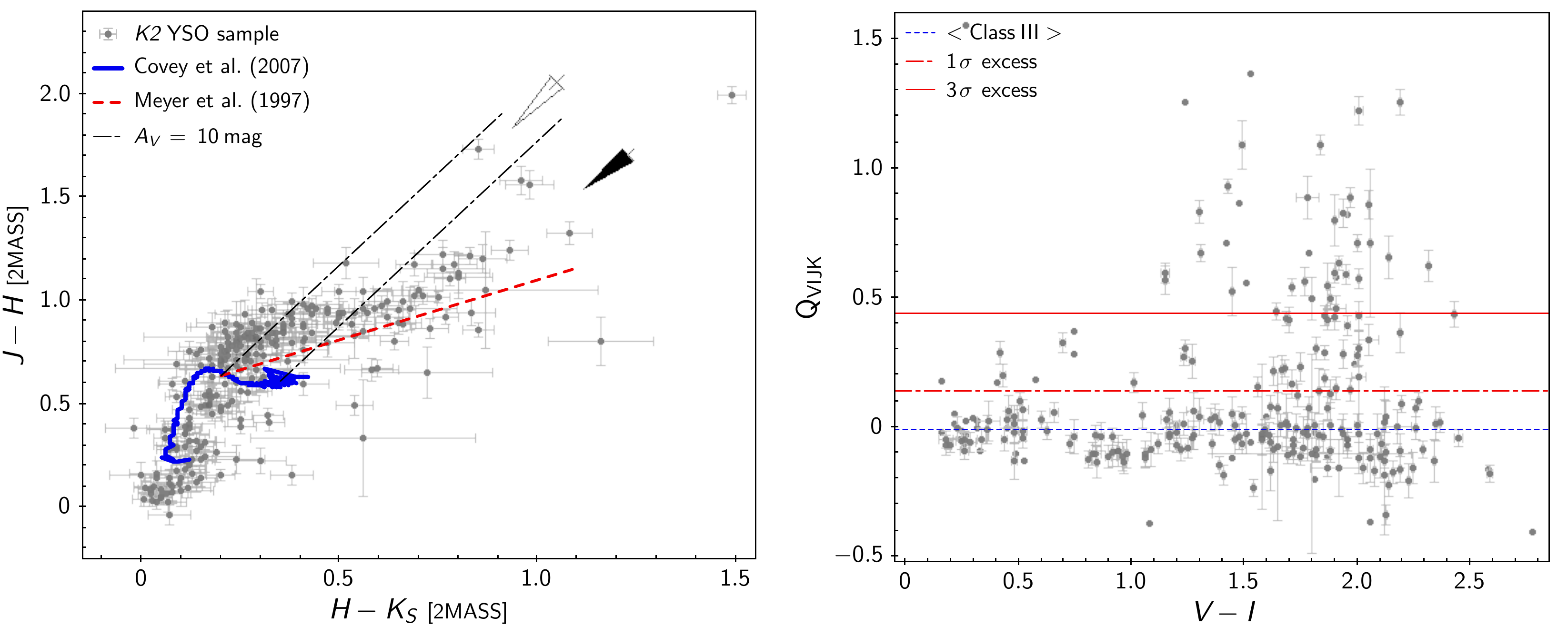}
\caption{{\it Left}: 2MASS $J,H,K$ color-color diagram for young stars in the Lagoon Nebula, monitored with {\it K2}. The blue curve traces the dwarf color locus tabulated in \citet{covey2007}. The dashed red line traces the {disk-bearing} YSO locus defined in \citet{meyer1997}. The {dash-dotted black} lines trace a 10~mag reddening vector starting from the beginning of the Meyer locus and from the end of the dwarf sequence. Objects above (or consistent with) the Meyer locus, and that fall {in the area indicated by the black arrow,} to the right of the second {black} line, were classified as {disk-bearing} sources; objects above (or consistent with) the Meyer locus, and that fall {in the area indicated by the white arrow,} between the first and the second {black} lines, were classified as disk candidates. {\it Right}: $Q_{VIJK}$ index calculated for young stars in the Lagoon Nebula, shown against their $V-I$ colors. The dotted {blue} line traces the typical $Q_{VIJK}$ value measured for Class~III sources in the sample, while the dash-dotted and the solid red lines mark a 1~$\sigma$ level and a 3~$\sigma$ level above the Class~III threshold, respectively. Objects that fall above the solid line were classified as {disk-bearing} sources, while those that fall between the dotted and the dash-dotted lines were classified as disk candidates according to this indicator.}
\label{fig:IR_exc_indicators}
\end{figure}

\begin{enumerate}
\item We used the 2MASS and UKIDSS data to build the $J,H,K$ color-color diagram, which is sensitive to dust in the innermost disk regions, at distances $<$0.1~AU from the central star. The resulting diagram is illustrated in Fig.\,\ref{fig:IR_exc_indicators} (left panel). Young stars with significant excess emission in $J,H,K$ bands are expected to distribute along a separate color locus from stars that exhibit purely photospheric emission, as discussed in \citet{meyer1997}.
\item We investigated the photometric properties of our stars in the wavelength range probed by {\it Spitzer}/IRAC filters (3.6--8.0~$\mu$m; also see Sect.\,\ref{sec:lc_IR_colors} below), sensitive to thermal emission from dust within the inner AU around solar-type stars, and verified which of the targets exhibit colors consistent with the Class~II locus defined by \citet{allen2004}.
\item We used the WISE photometry, when available, to build color-color diagrams in the W1 (3.4~$\mu$m), W2 (4.6~$\mu$m), and W3 (12~$\mu$m) passbands, and ascertained which sources exhibit color properties consistent with the Class~II locus defined by \citet{koenig2014}.
\item We calculated two reddening-free indices, $Q_{JHHK}$ and $Q_{VIJK}$, defined by \citet{damiani2006} as:
\begin{align} \label{eqn:Q_indices}
Q_{JHHK} &= (J-H) - \frac{A(J-H)}{A(H-K)}(H-K) \\
Q_{VIJK} &= (J-K) - \frac{A(J-K)}{A(V-I)}(V-I)
\end{align}
Such indices provide a measure of the near-IR color excess exhibited by disk-bearing stars with respect to disk-free stars in the region, computed after normalizing the cluster locus to the reddening direction on the color-color diagram. For each of the two indices, the typical photospheric value was defined as the average measured across {\it K2} targets in the Lagoon Nebula that were classified as Class~III sources upon a visual inspection of their SED. All objects that fell at a distance of over 3~$\sigma$ with respect to the typical Class~III value were then classified as disk sources, while objects that fell at a distance of 1-3~$\sigma$ from the typical Class~III locus were classified as disk candidates (Fig.\,\ref{fig:IR_exc_indicators}, right panel).
\end{enumerate}

Classifications from each indicator were then combined to assign a final disk status to each target: objects visually classified as Class~II SEDs that satisfy at least another IR excess indicator were retained as disked sources; objects visually classified as Class~III SEDs that do not exhibit any IR excess indicator were retained as disk-free sources; objects with discordant classifications or incomplete information, due to missing data, among the various indicators were flagged as potential disk candidates. Disked sources account for $\sim$31\% of our sample, disk-free sources for $\sim$50\%, and potential disk candidates for $\sim$19\%. 

\begin{deluxetable*}{c | c c c}
\centerwidetable
\tablecaption{Percentage of disk-bearing, disk-free, and disk candidate members of the Lagoon Nebula region, monitored with {\it K2}, as a function of their spectral class.}
\label{tab:Lagoon_SpT_disks}
\tablehead{\colhead{Spectral class} & \colhead{{Disk-bearing}} & \colhead{{Disk candidates}} & \colhead{Disk-free}}
\startdata
 B-type & 13\%$^{+0\%}_{-2\%}$ {\small $\mathit{\left[23\%\pm7\%\right]}$} &  {31\%$^{+4\%}_{-3\%}$ {\small $\mathit{\left[31\%\pm4\% \right]}$}} &  56\%$^{+5\%}_{-3\%}$ {\small $\mathit{\left[46\%\pm5\% \right]}$} \\
 A-type & 17\%$^{+3\%}_{-3\%}$ {\small $\mathit{\left[39\%\pm4\%\right]}$} & {23\%$^{+7\%}_{-3\%}$ {\small $\mathit{\left[15\%\pm1\% \right]}$}} &  60\%$^{+5\%}_{-7\%}$ {\small $\mathit{\left[46\%\pm4\% \right]}$} \\
 F-type & 12\%$^{+5\%}_{-2\%}$ {\small $\mathit{\left[32\%\pm7\%\right]}$} & {18\%$^{+5\%}_{-3\%}$ {\small $\mathit{\left[21\%\pm2\%\right]}$}} & 70\%$^{+3\%}_{-7\%}$ {\small $\mathit{\left[47\%\pm6\%\right]}$} \\
 G-type & 31\%$^{+5\%}_{-4\%}$ {\small $\mathit{\left[37\%\pm4\%\right]}$} & {20\%$^{+2\%}_{-3\%}$ {\small $\mathit{\left[19\%\pm2\%\right]}$}} & 49\%$^{+6\%}_{-5\%}$ {\small $\mathit{\left[44\%\pm4\%\right]}$} \\
 K-type & 36\%$^{+7\%}_{-4\%}$ {\small $\mathit{\left[56\%\pm3\%\right]}$} & {16\%$^{+3\%}_{-3\%}$ {\small $\mathit{\left[17\%\pm1\%\right]}$}} & 48\%$^{+4\%}_{-7\%}$ {\small $\mathit{\left[27\%\pm2\%\right]}$} \\
\enddata
\tablecomments{Values reported outside the square brackets correspond to the statistics measured from our targets with available SpT estimate from the analysis in Sect.\,\ref{sec:Av_SpT}, and the associated uncertainties are derived from the subsample of targets with no SpT estimate. The percentages reported in italic font between square brackets are projected statistics that account for the population of YSO candidates in the same magnitude range as our targets, but with no {\it K2} photometry (hence not included in this study).}
\end{deluxetable*}

Table~\ref{tab:Lagoon_SpT_disks} reports how the fractions of disk-bearing, disk-free, and disk candidate stars in our sample vary as a function of spectral class. The disk fraction does not appear to be uniform across the Lagoon Nebula population, but to be higher among later-type stars. In order to estimate an uncertainty on the measured disk fractions, and to evaluate the impact of sample incompleteness, we used the results of our SpT and disk classification analysis on the sample with {\it K2} data to derive the occurrence rates of each spectral class and disk class as a function of $R$-magnitudes for the entire {population of YSO candidates in the Lagoon Nebula region identified in our literature mining effort (see Sect.\,\ref{sec:membership}). More specifically, we selected, among the $\sim$3000 YSO candidates identified from the literature, those with Johnson-Cousins R-band magnitudes or SDSS $r$-band magnitudes (statistically calibrated to R-magnitudes) in the range $R\sim8-17.5$, within which our {\emph K2} targets are distributed. This selection produced a sample of $\sim$1100 YSO candidates. From our {\emph K2} targets, classified in SpT and disk status as described in Sects.\,\ref{sec:Av_SpT} and \ref{sec:disk_class}, respectively, we measured the statistical frequencies of different spectral classes and disk classes as a function of R, and estimated the associated errors from the subset of objects in our sample with no SpT estimate}. We then used these statistical frequencies and associated errors to randomly sample the population of {$\sim$1100} YSO candidates {selected upon their R-band or $r$-band magnitudes, as described above}, but with no {\it K2} data. At each sampling, we randomly selected magnitude-dependent SpT frequencies from the frequency distributions extracted from our targets, and determined a statistical estimate of disk-bearing, disk-free, and disk candidate objects for each spectral class across the entire population of YSO candidates. The average disk fractions obtained with this procedure in 250 random samplings, as well as the associated standard deviations, are also reported in Table~\ref{tab:Lagoon_SpT_disks}.

\section{The variability properties of young stars in the Lagoon Nebula} \label{sec:YSO_variability}

In this Section, we provide a characterization of the diverse variability behaviors exhibited by young stars in the Lagoon Nebula region. In Sect.\,\ref{sec:K2_lc_class}, we describe the classification scheme that we adopted to assign a label to each {\it K2} light curve based on the dominant morphological pattern of the observed flux variations. In Sect.\,\ref{sec:VST_variability}, we employ the simultaneous VST time series photometry to evaluate the amount of correlated variability exhibited by our targets at optical and UV wavelengths, as a function of optical brightness and of the {\it K2} variability type. In Sect.\,\ref{sec:var_class_colors}, we explore the color properties of our targets on multiple photometric diagrams to pinpoint the physical origin of the observed variability behaviors.

\subsection{{\it K2} light curve morphology classes} \label{sec:K2_lc_class}

To sort our sample of {\it K2} light curves into morphological classes, we adopted the metrics developed by \citet{cody2014} as part of the CSI~2264 project. The classification scheme is based on two indicators that probe two distinct properties of the light curves: the degree of periodicity or stochasticity of the luminosity pattern, and the degree of symmetry (i.e., the balance between brightening and fading trends) of the observed flux variations with respect to the typical luminosity state of the object. Eight main categories of variables are identified with this scheme:

\begin{itemize}
\item {\it periodic} (P), which exhibit repeated, sinusoidal-like flux patterns with little or no evolution in shape or amplitude from one cycle to the next;
\item {\it quasi-periodic symmetric} (QPS), which exhibit an overall periodic flux pattern, symmetric in amplitude below and above the typical luminosity level of the star, but with noticeable changes in shape and/or amplitude from one cycle to the next;
\item {\it stochastic} (S), which exhibit irregular flux variations with no apparent periodicity and no preference for brightening events over fading events or vice versa;
\item {\it bursters} (B), which exhibit irregular flux variations, prominently in the form of intense and short--lived brightening events on top of a flat or slowly--varying light curve continuum, occurring with no obvious periodicity but repeatedly over spans of days or weeks;
\item {\it quasi-periodic dippers} (QPD), which exhibit prominent fading events, in a recurring pattern with a detectable periodicity, although with changes in shape and depth from one cycle to the next;
\item {\it aperiodic dippers} (APD), which exhibit prominent fading events, possibly repeated but with no obvious periodicity along the time series;
\item {\it multi-periodic} (MP), which exhibit multiple periodicities (e.g., a beating pattern or pulsations);
\item {\it eclipsing binaries} (EB), which exhibit the characteristic photometric signatures of one companion transiting the other, superimposed on the out--of--eclipse individual variability patterns.
\end{itemize}

\begin{figure}
\centering
\includegraphics[width=\textwidth]{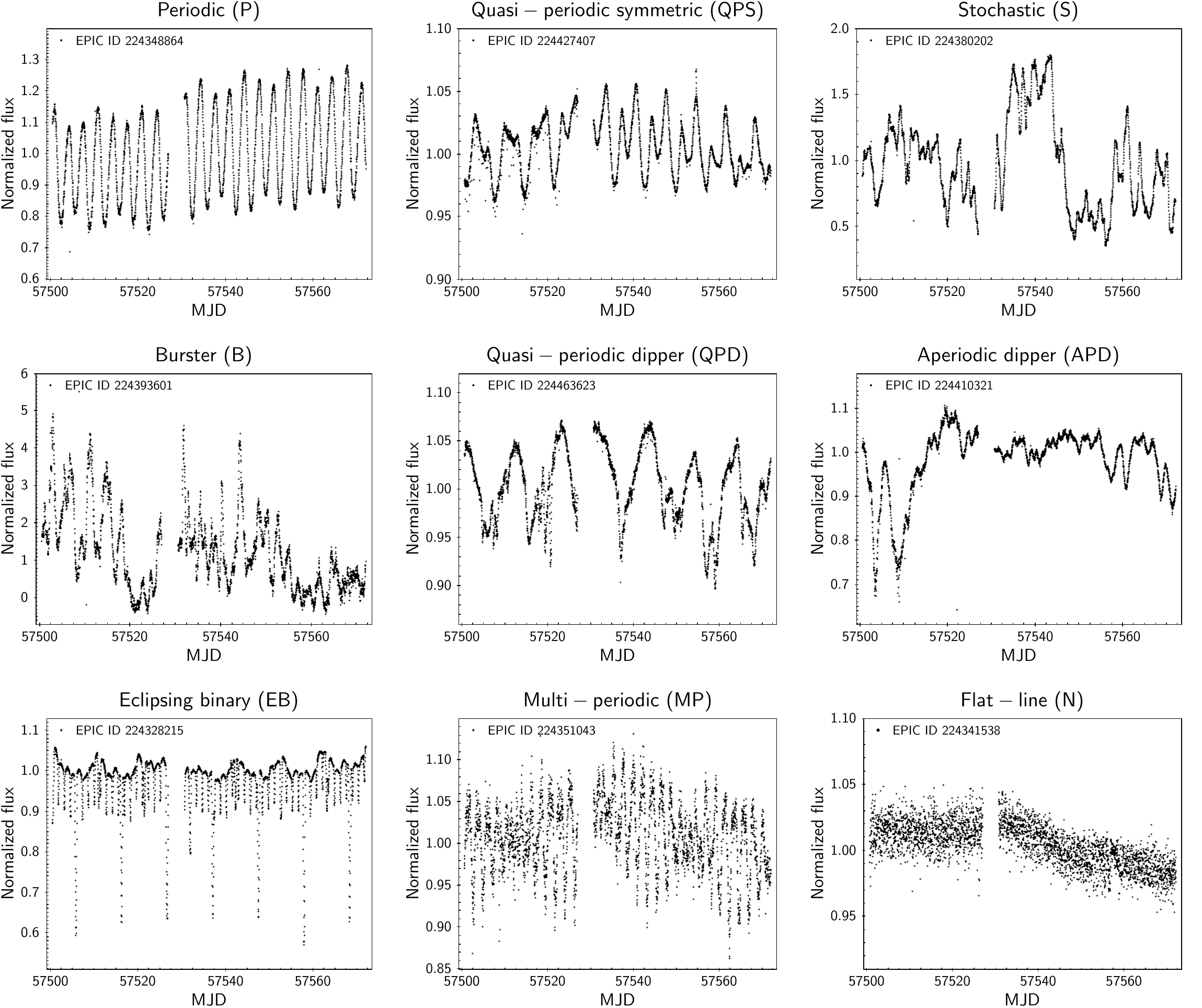}
\caption{Examples of the different types of variability identified among the {\it K2} light curves of Lagoon Nebula members.}
\label{fig:K2_lc_types}
\end{figure}

Additional groupings include {\it flat--line} or non-variable light curves (N), which exhibit no appreciable day-to-week variability patterns beyond statistical fluctuations, and {\it unclassifiable} patterns (U), which do not match any of the classes defined above. Representative examples of light curves belonging to different variability classes are illustrated in Fig.\,\ref{fig:K2_lc_types}.
{As can be seen, the intensity of flux variations is markedly different for distinct variable classes. Across our sample, regular variables like MP, QPS, and P sources, driven by magnetic starspots, possibly mixed with stable accretion spots, exhibit median peak-to-peak amplitudes (with $>$10\,$\sigma$ flux outliers removed) of order 12\%, 19\%, and 26\%, respectively, compared to the average flux level. While some underlying starspot variability is expected across all variable groups, this would constitute only a minor component in the most irregular photometric behaviors. Indeed, S variables exhibit median flux amplitudes $\sim$55\%, followed by QPD variables (78\%), APD variables (93\%), and B variables (125\%). Flat-line light curves, instead, merely exhibit median photometric fluctuations of order 3\% in flux.}

Table~\ref{tab:K2_class_disks} summarizes the occurrence rates of each variable class among our sample, as a function of disk status. While disked sources encompass all categories of variability patterns identified among {\it K2} targets, the most irregular and rapidly varying types of light curves (bursters, stochastic, dippers) are hardly found among non-disked stars, which supports an interpretation of those light curve patterns in terms of disk-related phenomena. Another inference from Table~\ref{tab:K2_class_disks} is that no stars with clear evidence of disks exhibit flat light curves, and very few of them exhibit unclear variability patterns, which indicates that the star--disk activity typically produces well recognizable photometric signatures. Disk candidate stars appear to be an intermediate group of objects between disked and non-disked sources, with mostly periodic or quasi-periodic variability behaviors, very limited cases of irregular variability, and a significant fraction of cases with unclassifiable variability patterns.

\begin{table}
\caption{Occurrence rates of different light curve types (described in Sect.\,\ref{sec:K2_lc_class}) across the entire population and as a function of disk status among Lagoon Nebula members. The percentage ranges reported in italics correspond to the $1\sigma$ confidence intervals ensuing from sample completeness levels in each of the three disk categories. These intervals were estimated by considering the fraction of missing objects and the prevalence of {disk-bearing, disk candidate, and disk-free} sources as a function of magnitude.}
\label{tab:K2_class_disks}
\centering
\begin{tabular}{c | c c c c}
\hline
{\it K2} class & Total & {Disk-bearing} & {Disk candidates} & {Disk-free} \\
 \hline
 P & 28.4\% & 17.4\% {\small $\mathit{\left[15\%-21\%\right]}$} & {35.8\% {\small $\mathit{\left[31\%-41\%\right]}$}} & 32.4\% {\small $\mathit{\left[30\%-35\%\right]}$} \\
 QPS & 27.7\% & 34.9\% {\small $\mathit{\left[31\%-39\%\right]}$} & {22.6\% {\small $\mathit{\left[19\%-27\%\right]}$}} & 25.2\% {\small $\mathit{\left[23\%-28\%\right]}$} \\
 S & 6.1\% & 16.3\% {\small $\mathit{\left[14\%-20\%\right]}$} & {1.9\% {\small $\mathit{\left[\lesssim 4\%\right]}$}} & 1.4\% {\small $\mathit{\left[1\%-2\%\right]}$} \\
 B & 1.8\% & 5.8\% {\small $\mathit{\left[4\%-8\%\right]}$} & {-- {\small $\mathit{\left[\lesssim 1\%\right]}$}} & -- {\small $\mathit{\left[< 1\%\right]}$} \\
 QPD & 3.6\% & 9.3\% {\small $\mathit{\left[7\%-12\%\right]}$} & {3.8\% {\small $\mathit{\left[3\%-6\%\right]}$}} & -- {\small $\mathit{\left[< 1\%\right]}$} \\
 APD & 2.5\% & 5.8\% {\small $\mathit{\left[4\%-8\%\right]}$} & {1.9\% {\small $\mathit{\left[\lesssim 4\%\right]}$}} & 0.7\% {\small $\mathit{\left[\lesssim 1\%\right]}$} \\
 MP & 14.4\% & 5.8\% {\small $\mathit{\left[4\%-8\%\right]}$} & {15.1\% {\small $\mathit{\left[12\%-19\%\right]}$}} & 19.4\% {\small $\mathit{\left[17\%-22\%\right]}$} \\
 EB & 2.5\% & 2.3\% {\small $\mathit{\left[2\%-4\%\right]}$} & {-- {\small $\mathit{\left[\lesssim 1\%\right]}$}} & 3.6\%  {\small $\mathit{\left[3\%-5\%\right]}$} \\
 N & 7.2\% & -- {\small $\mathit{\left[\lesssim 1\%\right]}$} & {5.7\% {\small $\mathit{\left[4\%-9\%\right]}$}} & 12.2\% {\small $\mathit{\left[11\%-14\%\right]}$} \\
 U & 5.8\% & 2.3\% {\small $\mathit{\left[2\%-4\%\right]}$} & {13.2\% {\small $\mathit{\left[10\%-17\%\right]}$}} & 5.0\%  {\small $\mathit{\left[4\%-6\%\right]}$} \\
 \hline
\end{tabular}
\end{table}

Table~\ref{tab:K2_class_SpT} summarizes the occurrence rates of different variability behaviors among Lagoon Nebula members of different spectral types. Between 60\% and 80\% of stars in each spectral class exhibit periodic, quasi-periodic, or multi-periodic variability patterns. This large fraction of stars with modulated light curves is consistent with earlier results from other {\it CoRoT} and {\it Kepler}/{\it K2} campaigns, including very young star clusters (\citealp[e.g.,][]{rebull2020} on Taurus; \citealp{rebull2018} on $\rho$~Ophiucus and Upper Scorpius; \citealp{venuti2017} on NGC~2264) and somewhat older, nearby clusters (\citealp[e.g.,][]{rebull2016} on the $\sim$125~Myr-old Pleiades; \citealp{rebull2017} on the $\sim$790~Myr-old Praesepe). We note, however, that other surveys have suggested significantly lower fractions of periodic variables in similarly aged clusters \citep[e.g.,][on the $\sim$164~Myr-old NGC~2301]{howell2005}.

Excluding B-type stars (a large fraction of which exhibit unclassifiable variability behaviors, while the remaining majority show regular modulated variability), a few trends can be inferred from Table~\ref{tab:K2_class_SpT}:
\begin{itemize}
\item the fraction of non-variable stars (which exhibit flat-line light curves) is significantly larger among earlier-type (A, F) stars than among later-type (G, K) stars;
\item while a small fraction of stochastic variables are also found among early-type stars, dipping behaviors are only observed among later-type stars;
\item multi-periodic variables occur more frequently among earlier-type stars than among later-type stars, while the opposite is true for single-periodicity variables (strictly periodic or quasi-periodic symmetric);
\item bursting behaviors appear to be uncommon and concentrated predominantly among higher-mass (A-type, F-type) stars, although the latter inference may be affected by our limited statistics, as discussed in Sect.\,\ref{sec:var_mass_dependence}. 
\end{itemize}

\begin{table}
\caption{Occurrence rates of different variability types among Lagoon Nebula members as a function of spectral class. Uncertainties are estimated from the magnitude distribution of objects in our sample with no SpT estimate, and from the magnitude-dependent distribution of different spectral classes in our sample. The ranges between square brackets correspond to the 1$\sigma$ confidence intervals calculated by taking into account what fraction of the estimated YSO population per spectral class is included in our sample.}
\label{tab:K2_class_SpT}
\centering
\begin{tabular}{c | r @{\hspace{1\tabcolsep}} l r @{\hspace{1\tabcolsep}} l r @{\hspace{1\tabcolsep}} l r @{\hspace{1\tabcolsep}} l r @{\hspace{1\tabcolsep}} l}
\hline
\multirow{2}{*}{{\it K2} class} & \multicolumn{2}{c}{B-stars (\%)} & \multicolumn{2}{c}{A-stars (\%)} & \multicolumn{2}{c}{F-stars (\%)} & \multicolumn{2}{c}{G-stars (\%)} & \multicolumn{2}{c}{K-stars (\%)} \\
 & \multicolumn{2}{c}{\small {[16 YSOs]}} & \multicolumn{2}{c}{\small {[30 YSOs]}} & \multicolumn{2}{c}{\small {[17 YSOs]}} & \multicolumn{2}{c}{\small {[39 YSOs]}} & \multicolumn{2}{c}{\small {[124 YSOs]}} \\
 \hline
P & 25.0$^{+4.4}$ & {\small $\mathit{\left[16-38\right]}$} & 16.7$_{-2.0}$ & {\small $\mathit{\left[12-24\right]}$} & 22.2$^{+4.1}_{-3.2}$ & {\small $\mathit{\left[16-32\right]}$} & 23.1$^{+3.7}_{-2.2}$ & {\small $\mathit{\left[19-29\right]}$} & 38.1$^{+2.4}_{-6.3}$ & {\small $\mathit{\left[35-42\right]}$}\\
 QPS & 25.0$_{-1.5}$ & {\small $\mathit{\left[16-38\right]}$} & 16.7$^{+2.7}_{-1.5}$ & {\small $\mathit{\left[12-24\right]}$} & 11.1$^{+4.7}_{-1.6}$ & {\small $\mathit{\left[8-19\right]}$} & 35.9$^{+1.6}_{-4.1}$ & {\small $\mathit{\left[30-43\right]}$} & 28.6$^{+4.2}_{-4.3}$ & {\small $\mathit{\left[26-32\right]}$}\\
 S & \multicolumn{1}{c}{--} & {\small $\mathit{\left[\lesssim 6\right]}$} & 3.3$_{-0.4}$ & {\small $\mathit{\left[2-8\right]}$} & 0.0$^{+5.3}$ & {\small $\mathit{\left[\lesssim 5\right]}$} & 5.1$^{+2.4}_{-0.6}$ & {\small $\mathit{\left[4-9\right]}$} & 5.6$^{+3.6}_{-2.0}$ & {\small $\mathit{\left[4-7\right]}$}\\
 B & \multicolumn{1}{c}{--} & {\small $\mathit{\left[\lesssim 6\right]}$} & 3.3$_{-0.4}$ & {\small $\mathit{\left[2-8\right]}$} & 5.6$_{-1.1}$ & {\small $\mathit{\left[3-12\right]}$} & \multicolumn{1}{c}{--} & {\small $\mathit{\left[\lesssim 2\right]}$} & 0.0$^{+1.6}$ & {\small $\mathit{\left[< 1\right]}$} \\
 QPD & \multicolumn{1}{c}{--} & {\small $\mathit{\left[\lesssim 6\right]}$} & \multicolumn{1}{c}{--} & {\small $\mathit{\left[\lesssim 3\right]}$} & \multicolumn{1}{c}{--} & {\small $\mathit{\left[\lesssim 5\right]}$} & 5.1$_{-0.7}$ & {\small $\mathit{\left[4-9\right]}$} & 5.6$^{+0.7}_{-1.1}$ & {\small $\mathit{\left[4-7\right]}$} \\
 APD & \multicolumn{1}{c}{--} & {\small $\mathit{\left[\lesssim 6\right]}$} & \multicolumn{1}{c}{--} & {\small $\mathit{\left[\lesssim 3\right]}$} & \multicolumn{1}{c}{--} & {\small $\mathit{\left[\lesssim 5\right]}$} & 2.6$_{-0.4}$ & {\small $\mathit{\left[2-6\right]}$} & 2.4$^{+2.3}_{-0.4}$ & {\small $\mathit{\left[2-4\right]}$}\\
 MP & 12.5$_{-0.7}$ & {\small $\mathit{\left[7-23\right]}$} & 30.0$^{+2.3}_{-2.7}$ & {\small $\mathit{\left[23-38\right]}$} & 27.8$_{-5.1}$ & {\small $\mathit{\left[20-38\right]}$} & 15.4$_{-2.1}$ & {\small $\mathit{\left[12-21\right]}$} & 14.3$^{+0.7}_{-2.7}$ & {\small $\mathit{\left[12-17\right]}$}\\
 EB & \multicolumn{1}{c}{--} & {\small $\mathit{\left[\lesssim 6\right]}$} & 3.3$_{-0.4}$ & {\small $\mathit{\left[2-8\right]}$} & 5.6$_{-1.1}$ & {\small $\mathit{\left[3-12\right]}$} & 2.6$_{-0.4}$ & {\small $\mathit{\left[2-6\right]}$} & 0.8$^{+0.8}_{-0.2}$ & {\small $\mathit{\left[\lesssim 1\right]}$}\\
 L & \multicolumn{1}{c}{--} & {\small $\mathit{\left[\lesssim 6\right]}$} & \multicolumn{1}{c}{--} & {\small $\mathit{\left[\lesssim 3\right]}$} & \multicolumn{1}{c}{--} & {\small $\mathit{\left[\lesssim 5\right]}$} & \multicolumn{1}{c}{--} & {\small $\mathit{\left[\lesssim 2\right]}$} & \multicolumn{1}{c}{--} & {\small $\mathit{\left[< 1\right]}$} \\
 N & \multicolumn{1}{c}{--} & {\small $\mathit{\left[\lesssim 6\right]}$} & 20.0$^{+2.6}_{-1.8}$ & {\small $\mathit{\left[14-28\right]}$} & 22.2$^{+4.1}_{-3.2}$ & {\small $\mathit{\left[16-32\right]}$} & 5.1$^{+4.7}_{-0.4}$ & {\small $\mathit{\left[4-9\right]}$} & 1.6$^{+1.5}_{-0.3}$ & {\small $\mathit{\left[1-3\right]}$}\\
 U & 37.5$_{-2.2}$ & {\small $\mathit{\left[27-50\right]}$} & 6.7$^{+3.0}_{-0.6}$ & {\small $\mathit{\left[4-12\right]}$} & 5.6$_{-1.1}$ & {\small $\mathit{\left[3-12\right]}$} & 5.1$_{-0.7}$ & {\small $\mathit{\left[4-9\right]}$} & 3.2$^{+1.5}_{-0.6}$ & {\small $\mathit{\left[3-5\right]}$}\\
 \hline
\end{tabular}
\end{table}

\subsection{Multiwavelength correlated variability from VST/OmegaCAM data} \label{sec:VST_variability}

To assess the amount of intrinsic variability, beyond the photometric noise level, among Lagoon Nebula members, we implemented the variability index $\mathcal{J}$ defined in \citet{stetson1996}. The $\mathcal{J}$ index measures the amount of correlated variability observed for a given star at two different wavelengths $m$ and $m'$:
\begin{equation} \label{eqn:Jindex}
\mathcal{J} = \frac{\sum_{k=1}^{n} w_k \sgn{(P_k)} \sqrt{|P_k|}}{\sum_{k=1}^{n} w_k},
\end{equation}
where $n$ is the number of time-ordered observations, $P_k = \frac{n}{n-1} \left(\frac{m_k - \overline{m}}{\sigma_m}\right)\left(\frac{m'_k - \overline{m}'}{\sigma_{m'}}\right)$, {${\sgn{(P_k)}}$ takes value +1 if $P_k$ is positive and -1 if $P_k$ is negative,} $m_k$ and $m'_k$ are the magnitudes measured during the $k^{th}$ observing epoch, $\overline{m}$ and $\overline{m}'$ are the average magnitudes measured across the time series, $\sigma_m$ and $\sigma_{m'}$ are the photometric uncertainties, and $w_k$ is a weight assigned to each pair of observations $m_k$ and $m'_k$ to take into account, for instance, the exact time lag $\Delta t_k$ between observations in the $k^{th}$ pair compared to the average $\Delta t$ measured across all paired observations \citep[e.g.,][]{fruth2012, venuti2015}. If the physical drivers of the variability observed at different wavelengths are interconnected, magnitude variations observed simultaneously in different filters are expected to be correlated, yielding a non-zero sum for $\mathcal{J}$. Conversely, if the photometric fluctuations in the light curve are primarily driven by noise, no correlation is expected between the magnitude variations measured in different filters, and the $\mathcal{J}$ index will converge to zero.

\subsubsection{Implementation of the $\mathcal{J}$ index variability diagnostics}

\begin{figure}[t]
\centering
\includegraphics[width=\textwidth]{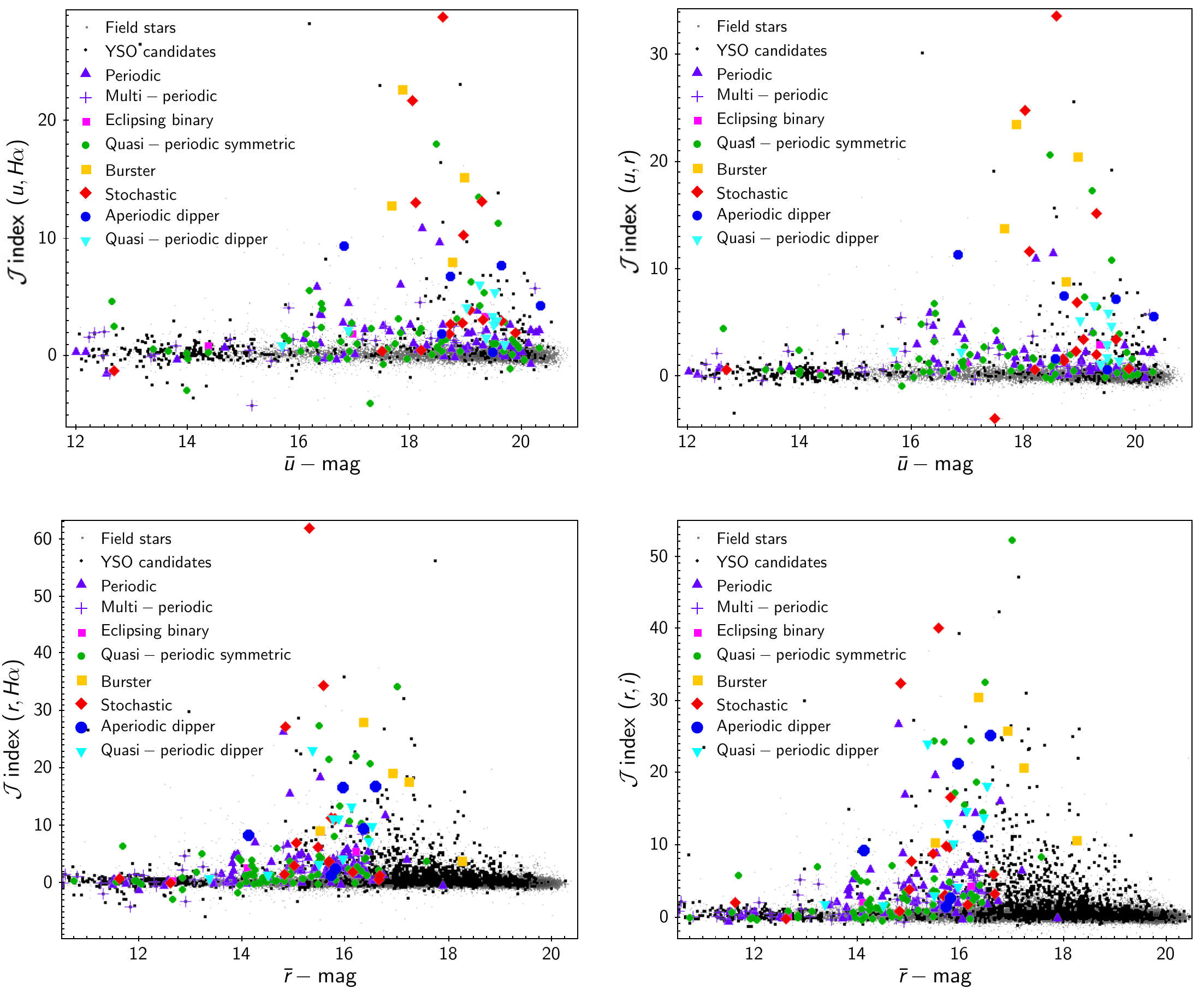}
\caption{Values of $\mathcal{J}$ index calculated to probe correlated variability in $u$-band and $H\alpha$-band (top left), $u$-band and $r$-band (top right), $r$-band and $H\alpha$-band (bottom left), and $r$-band and $i$-band (bottom right). In each panel, the same color and symbol code is used to distinguish among different stellar groups: field stars (small gray dots), YSO candidates (larger black dots), and various classes of variability identified among {\it K2} targets (P $\rightarrow$ purple upward triangles, MP $\rightarrow$ purple crosses, EB $\rightarrow$ small fuchsia squares, QPS $\rightarrow$ green-filled circles, B $\rightarrow$ large yellow squares, S $\rightarrow$ red diamonds, APD $\rightarrow$ large blue-filled circles, QPD $\rightarrow$ cyan downward triangles). YSOs exhibit on average larger amounts of correlated variability than field stars, and irregular {\it K2} variables exhibit more intense flux variations (translating to higher $\mathcal{J}$) than regular variables.}
\label{fig:Stetson_J}
\end{figure}

Equation~\ref{eqn:Jindex} was applied to compute the $\mathcal{J}$ index between simultaneous variations in the $u$-band, $r$-band, and $H\alpha$-band (sensitive to accretion signatures), and between $r$-band and $i$-band (sensitive to photospheric emission and surface spot modulation). Results are illustrated in Fig.\,\ref{fig:Stetson_J}.
As expected, field stars are distributed along a narrow sequence close to $\mathcal{J}=0$, albeit with a small wavelength dependence possibly due to the different levels of photometry accuracy that were attained during long exposures and short exposures to recover the faint and the bright components of the population. A significant dispersion in variability properties is observed among YSO candidates, with a fraction of objects distributed along the sequence of field stars, and another fraction located at larger $\mathcal{J}$ than field stars, which indicates a higher level of intrinsic variability for PMS stars than for main sequence objects. Class~III (disk-free) YSO candidates exhibit an average $\mathcal{J}$ index ten times larger than that measured for field stars in redder filters (i.e., $\mathcal{J}$\,($r,H\alpha$) and $\mathcal{J}$\,($r,i$)), and six times larger in the $u$-band (i.e., $\mathcal{J}$\,($u,H\alpha$) and $\mathcal{J}$\,($u,r$)), affected by larger photometric uncertainties. Class~II (disk-bearing) YSO candidates, instead, exhibit average $\mathcal{J}$\,($r,H\alpha$) and $\mathcal{J}$\,($r,i$) 40 times larger, and average $\mathcal{J}$\,($u,H\alpha$) and $\mathcal{J}$\,($u,r$) $>$20 times larger, than those measured for field stars.

\subsubsection{Statistical significance of the $\mathcal{J}$ index trends}

To assess the significance of the distinct trends observed for field stars and YSO candidates in Fig.\,\ref{fig:Stetson_J}, we applied the Energy Test described in \citet{aslan2005}. Namely, we treated the two-dimensional distributions of the two populations (field stars vs. YSO candidates) on the (mag,~$\mathcal{J}$) diagrams as two systems of charges, one carrying positive charge, and the other carrying negative charge. Each individual point in a given system was assigned a charge of magnitude $1/n$, where $n$ is the number of objects in the corresponding distribution, so that each system carries a total charge of unitary magnitude. The test statistic $\Phi$ is defined as the sum of three terms: the potential energy of the two systems of charges, $\Phi'$ and $\Phi''$, and the interaction energy between the two systems of charges, $\Phi^{int}$. Each term is defined as follows:
\begin{itemize}
\item $\Phi' = \frac{1}{(n')^2}\sum_{i=1}^{n'}\sum_{j=i+1}^{n'}f(|{\bf x'}_i - {\bf x'}_j|)$;
\item $\Phi'' = \frac{1}{(n'')^2}\sum_{i=1}^{n''}\sum_{j=i+1}^{n''}f(|{\bf x''}_i - {\bf x''}_j|)$;
\item $\Phi^{int} = -\frac{1}{n'n''}\sum_{i=1}^{n'}\sum_{j=1}^{n''}f(|{\bf x'}_i - {\bf x''}_j|)$.
\end{itemize}
In the above definitions, $|{\bf x}_i - {\bf x}_j|$ is the Euclidean distance $d$ between points ${\bf x}_i$ and ${\bf x}_j$ on the (mag, $\mathcal{J}$) diagram, and $f(|{\bf x}_i - {\bf x}_j|)$ is a function of $d$, defined in \citet{aslan2005} as $f(d) = -\ln{d}$. To prevent the variations along one axis from prevailing numerically over the variations along the other axis, we renormalized all coordinates as $x'_1 = (x_1 - \overline{x_1})/\sigma_{x_1}$ and $x'_2 = (x_2 - \overline{x_2})/\sigma_{x_2}$, where $x_1$ and $x_2$ are the datapoint coordinates on the (mag, $\mathcal{J}$) diagram, $\overline{x_1}$ and $\overline{x_2}$ are the average coordinates measured across the sample, and $\sigma_{x_1}$ and $\sigma_{x_2}$ are the standard deviations measured around the average coordinates across the sample. For sufficiently large numbers of objects, the total potential energy $\Phi$ is expected to reach its minimum value if the two systems of charges are similarly distributed on the plane. Therefore, by measuring how different the measured $\Phi$ value is compared to the minimum $\Phi$ expectation, we can assess whether the null hypothesis of the two populations being drawn from the same distribution can be rejected to a certain significance. 

To build a statistical distribution for the expected minimum $\Phi$ under the null hypothesis, we implemented a permutation test as described in \citet{efron1993}. Namely, we merged the two populations (field stars and YSOs) into a single sample of $n'+n''$ objects. Then, we randomly picked $n'$ objects from the merged sample, with no replacements and no repetitions, to represent the first test population; the remaining $n''$ objects comprised the second test population. By construction, these two test populations derive from the same statistical distribution, and therefore the $\Phi$ statistic measured for them is a realization of the null hypothesis. We repeated this procedure 10\,000 times, and compared the value of $\Phi$ measured for our science populations to the average $\bar{\Phi}$ and its dispersion $\sigma_\Phi$ simulated under the null hypothesis. The test results show clear evidence of a significant difference between the (mag,~$\mathcal{J}$) distribution of field stars and of YSOs when $u+r$, $u+H\alpha$, $r+H\alpha$, and $r+i$ bands are considered. In each of the four cases, no occurrence of $\Phi$ larger than that measured for the science samples was obtained after 10\,000 permutation resamples, which indicates that the null hypothesis can be rejected to a $p$-level~$\lesssim 0.0001$.

\subsubsection{$\mathcal{J}$ index vs. spectral class and {\it K2} variability} \label{sec:Jindex_spt}

\begin{figure}
\centering
\includegraphics[width=\textwidth]{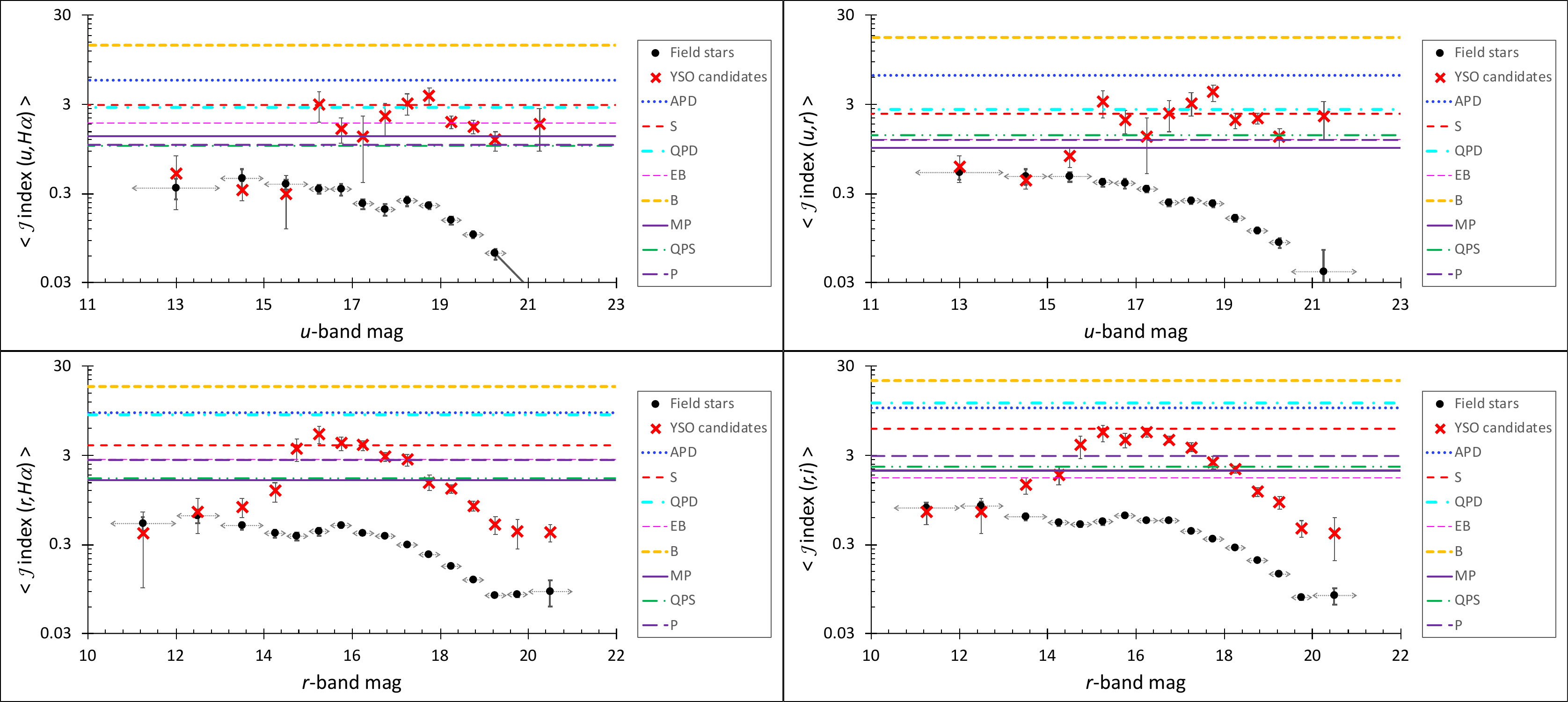}
\caption{Average values of $\mathcal{J}$ index measured, as a function of $u$-band magnitude (top) and $r$-band magnitude (bottom), between $u$ and $H\alpha$ bands (top left), $u$ and $r$ bands (top right), $r$ and $H\alpha$ bands (bottom left), and $r$ and $i$ bands (bottom right). Black dots correspond to the average measurements derived for field stars, and red crosses identify those derived for YSO candidates in the same magnitude bins. The double arrows associated with the field star measurements along the x-axis mark the extent of the magnitude bins in which the average measurements were obtained. $< \mathcal{J} \mbox{ index} >$ estimates, for both the field population and the YSOs population, were obtained with the bootstrap method. The associated error bars correspond to the standard deviation of the average $\mathcal{J}$ values measured in each of the 100\,000 bootstrap--resampled populations used. Horizontal lines mark the median $\mathcal{J}$ indices computed for different classes of {\it K2} variables, coded in color as in Fig.\,\ref{fig:Stetson_J}, and in dash type as shown in the side legend.}
\label{fig:average_Jind}
\end{figure}

To perform a more quantitative comparison between the variability properties of YSO candidates and field stars, we sorted our sample into magnitude bins and, for each of them, we computed the typical $\mathcal{J}$ index for both stellar groups. In order to account for the large scatter in value observed at each magnitude on Fig.\,\ref{fig:Stetson_J}, we applied the bootstrap method, and created 100\,000 resampled populations for both field stars and YSOs. For each resample, we computed the average $\mathcal{\bar{J}}$ index in each magnitude bin; we then extracted the average $<\mathcal{J}>$ as the mean of all $\mathcal{\bar{J}}$ computed in each magnitude bin, while their standard deviation was adopted as the uncertainty associated with the $<\mathcal{J}>$ index measurement. Figure \ref{fig:average_Jind} illustrates the results of this analysis as a function of magnitude. 
Field stars are expected to trace the level of photometric noise across the magnitude range. In all panels of Fig.\,\ref{fig:average_Jind}, the brightest YSO members (down to $\sim$early-G spectral types) do not exhibit statistical variability above the field population level. Conversely, for fainter magnitudes and later spectral types, the sequence of YSO candidates is clearly distinct from the level traced by field stars, at least down to the faintest objects ($r \geq 19$, well beyond the magnitude limit attained with {\it K2}), where photometric uncertainties are considerably larger than for brighter sources (see Fig.\,\ref{fig:VST_ur_rms}).

As observed in Figs.\,\ref{fig:Stetson_J} and \ref{fig:average_Jind}, different levels of variability are statistically associated with different classes of {\it K2} variables. Regular variability patterns (P, QPS, MP, EB) on all diagrams are associated with smaller $\mathcal{J}$ indices than irregular and rapidly-evolving variability patterns (B, S, QPD, APD). {This trend reflects the fact that B, S, QPD, and APD variables display larger amplitudes of variability (hence larger differences between instantaneous magnitudes and average magnitudes in the $\mathcal{J}$ index definition) than modulated variables.} Bursters and aperiodic dippers, in the order, exhibit the largest amounts of correlated variability in all diagrams, followed by quasi-periodic dippers and stochastic stars. {This progression mirrors the comparative intensities of flux variations that are associated on average with each irregular class, both in the {\it K2} time series (see Sect.\,\ref{sec:K2_lc_class}) and in all VST/OmegaCAM filters except the $u$-band, where stochastic variables typically display larger amplitudes of variability than dipper variables.} Stochastic stars constitute also the variable class that exhibits the largest case-by-case dispersion in measured $\mathcal{J}$ index, as can be seen on Fig.\,\ref{fig:Stetson_J}.

\subsection{Variability behaviors and color properties} \label{sec:var_class_colors}
In this Section, we illustrate the color properties of different YSO variables on UV-to-IR photometric diagrams, to identify the leading causes of different variability behaviors, with a particular focus on disk-related phenomena.

\subsubsection{Light curve morphology and IR colors} \label{sec:lc_IR_colors}

\begin{figure}
\centering
\includegraphics[width=\textwidth]{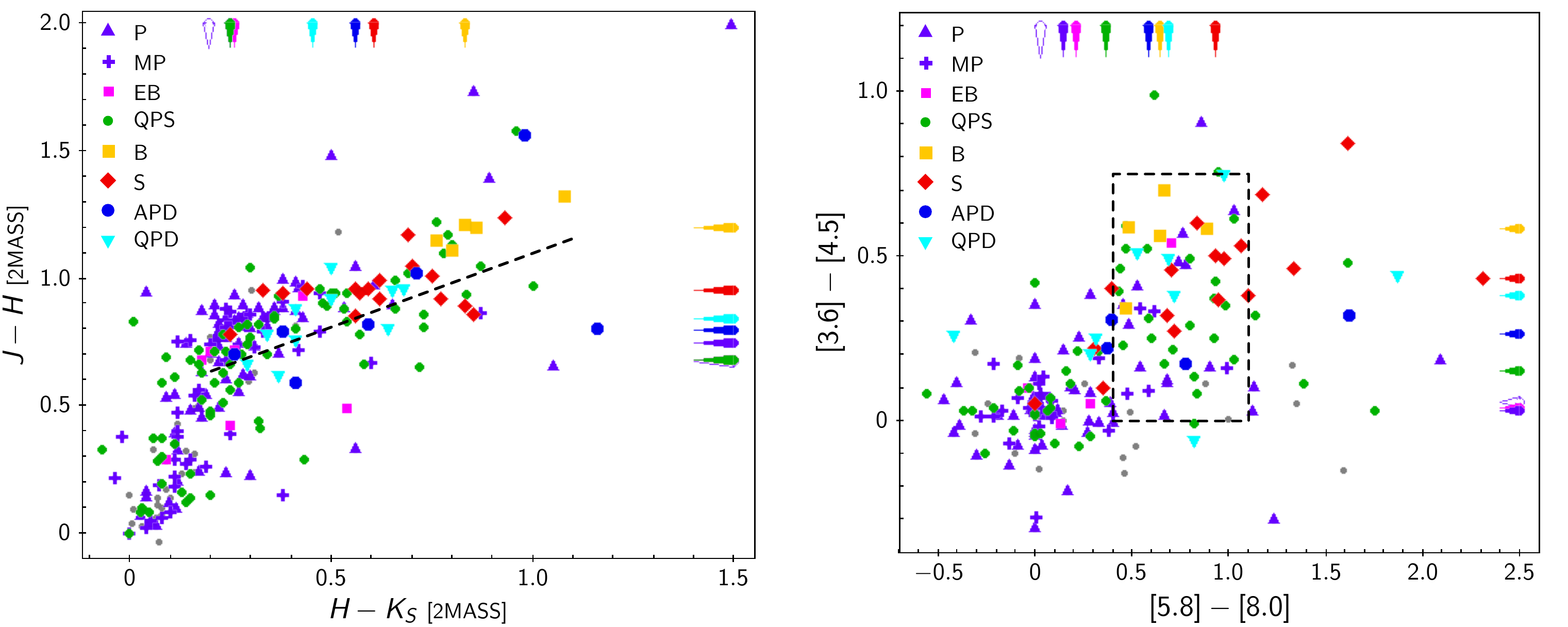}
\caption{{\it Left}: Near-IR color properties of {\it K2} Lagoon Nebula members in 2MASS $J,H,K_S$ filters. Different colors and symbols correspond to distinct classes of {\it K2} variables, following the same notation introduced in Fig.\,\ref{fig:Stetson_J} and summarized in the diagram legend. Markings along the upper and right-hand axes indicate, respectively, the median $H-K_S$ and $J-H$ colors measured for the different variable classes {(a filled purple arrow is used to indicate the median colors for P variables, while a contoured arrow is used for MP variables). The color locus for disk-bearing sources defined in \citet{meyer1997} is traced as a dashed line on the diagram}. {\it Right}: Mid-IR colors measured for {\it K2} Lagoon Nebula members in {\it Spitzer}/IRAC filters. Different symbols correspond to different variable classes, similarly to what shown in the left panel. The median colors measured for each class of {\it K2} variables are marked along the upper and right-hand diagram axes. The dashed box indicates the IRAC locus dominated by Class~II objects as defined in \citet{allen2004}.}
\label{fig:NIR_MIR_colors_K2}
\end{figure}

YSO colors at IR wavelengths probe the thermal emission from circumstellar material at different radii from the central star. Figure~\ref{fig:NIR_MIR_colors_K2} illustrates the color properties of Lagoon Nebula members, sorted according to their {\it K2} light curve morphology, from the $J$-band ($\sim$1.2~$\mu$m) to 8~$\mu$m wavelengths. These wavelengths trace reprocessed emission from dust located at distances between $\lesssim$0.1~AU (near-IR) and $\sim$5~AU (mid-IR) around solar-type stars \citep{williams2011}. 

The observed color properties suggest distinct physical origins for regular and irregular variability behaviors. Conspicuous amounts of material in the inner disk regions appear to be characteristic of burster-like, stochastic-like, and dipper-like variables, therefore confirming that these behaviors arise from disk-related phenomena. Modulated behaviors like periodic and quasi-periodic variables, instead, are typically associated with color properties consistent with stellar photospheric emission \citep[e.g.,][]{covey2007}. The typical IR excess emission exhibited by burster stars is larger than, in the order, that associated with stochastic stars and dipper stars at $\sim$1.2~$\mu$m to 4.5~$\mu$m wavelengths. Stochastic stars, on the other hand, appear to exhibit typically redder $5.8\,\mu$m\,--\,$8.0\,\mu$m colors than bursters and dippers. This may indicate that stochastic variability is driven by time-variable accretion streams that develop beyond the innermost disk regions; however, we do not have sufficient data for our sample from the far-IR {\it Spitzer}/Multiband Imaging Photometer (MIPS) or WISE surveys to assess whether this trend extends to longer ($\sim20\,\mu$m) wavelengths.

\subsubsection{Light curve morphology and UV photometry}

Mass accretion from the inner disk onto the central star is a main ingredient in the dynamics of star--disk interaction in YSOs, and UV excess emission represents its most prominent observational signature, as discussed in Sect.\,\ref{sec:intro}. While broadband spectroscopy \citep[e.g.,][]{manara2013} provides the most accurate determination of accretion activity above the emission level of the stellar photosphere, U-band (or $u$-band) photometry has long been shown to be an efficient, reliable proxy to the total accretion luminosity \citep{gullbring1998}, and it has been used in the literature to map the accretion properties of numerous pre--main sequence populations (e.g., \citealp{sicilia-aguilar2010}; \citealp{rigliaco2011}; \citealp{venuti2014}).

\begin{figure}
\centering
\includegraphics[width=\textwidth]{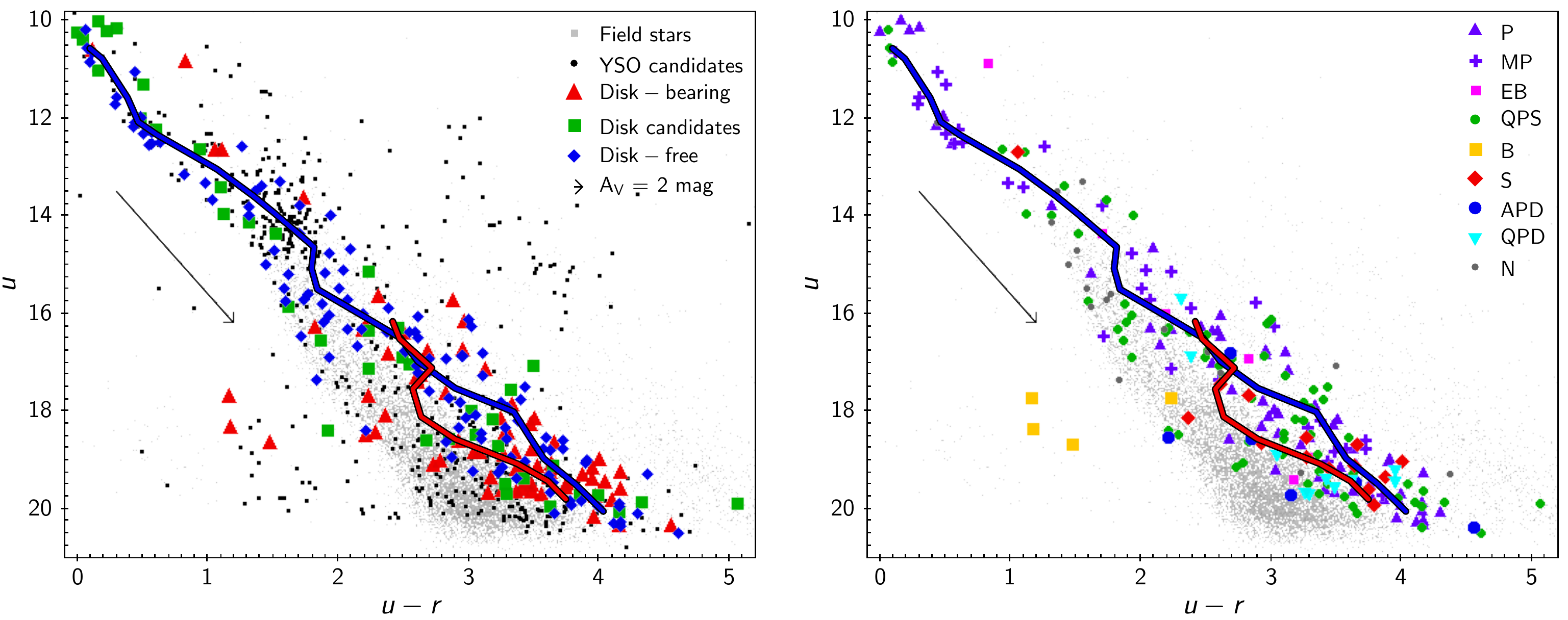}
\caption{{\it Left}: ($u-r$, $u$) color-magnitude diagram for stars in the Lagoon Nebula field surveyed with VST/OmegaCAM. Gray dots correspond to field stars, black dots to YSO candidates in the field. Red triangles, green squares, and blue diamonds correspond respectively to disk--bearing, disk candidate, and disk--free Lagoon Nebula members monitored with {\it K2}. The blue curve and the red curve trace respectively the average $u-r$ colors measured as a function of $u$-band magnitude for disk--free (Class~III) and disk--bearing (Class~II) objects. The reddening vector on the left of the diagram traces the photometric shift induced by a 2~mag extinction according to a reddening law with $R_V = 5$. {\it Right}: Same diagram as in the left panel, with ${\it K2}$ Lagoon Nebula YSOs sorted according to their light curve morphology. Colors and symbols are associated with different variability classes following the same convention as in Fig.\,\ref{fig:Stetson_J}. Non--variable stars (i.e., flat--line light curves) are indicated as charcoal-gray dots.}
\label{fig:ur_u_disk_K2}
\end{figure}

Figure~\ref{fig:ur_u_disk_K2} illustrates the $u-r$ colors of young stars in the Lagoon Nebula in relation to their $u$-band magnitudes. Confirmed cluster members monitored with {\it K2} form a distinct, albeit scattered, sequence, located above the distribution of field stars (i.e., at brighter magnitudes) at any given $u-r$ interval. While the distribution of disk--bearing and disk--free stars are largely overlapping (Fig.\,\ref{fig:ur_u_disk_K2}, left panel), we did ascertain distinct trends in the average $u-r$ colors measured for disk-free and disk-bearing stars as a function of $u$-mag. Namely, we calculated a moving average of the observed $u-r$ for different YSO groups by sampling the $u$-magnitude range in 1~mag--wide bins with a step of 0.5~mag. The derived average trends of $u$ vs. $u-r$ for disk--free and disk--bearing objects are shown respectively as a blue curve and a red curve on Fig.\,\ref{fig:ur_u_disk_K2} (left). The average $u-r$ measured for disk--free stars in a given $u$-mag bin tends to be larger than that measured, in the same magnitude range, for disk--bearing stars at $u \gtrsim 17.5$, where Class~II YSOs are most represented. Disk candidate YSOs, as classified in Sect.\,\ref{sec:disk_class}, follow a similar color distribution to that of disk--bearing YSOs, as the majority of them lies to the left of the average sequence for disk--free stars (i.e., at lower $u-r$ values). The lower-on-average $u-r$ colors measured for disk--bearing and potential disk--bearing objects, as opposed to disk--free objects, suggests that many of these sources exhibit some excess emission in the $u$-band, indicative of ongoing accretion activity. 

The right panel of Fig.\,\ref{fig:ur_u_disk_K2} shows how different types of {\it K2} variables compare in terms of their $u-r$ properties. Strictly periodic variables tend to follow closely the average ($u-r, u$) sequence of disk--free stars, as also observed for quasi--periodic variables, albeit with a larger scatter around the average sequence. This suggests that a modulated variability behavior is driven primarily by photospheric features. Dipper stars, on the other hand, tend to be located to the left of the average disk--free color sequence, and burster stars tend to exhibit $u-r$ color excesses significantly larger than the average measured for typical disk--bearing YSOs. This result confirms previous findings that a bursting behavior is associated with high accretion levels (\citealp{stauffer2014}; \citealp{venuti2014}; \citealp{cody2017}). Stochastic stars also appear to be primarily located at bluer colors than the average photospheric behavior, but, contrary to burster stars, they do not stand out with respect to other disk--bearing objects.

\subsubsection{Light curve morphology and H$\alpha$ photometry} \label{sec:lc_Ha_emission}

H$\alpha$ line emission is another widely used indicator of ongoing disk accretion onto young stars. H$\alpha$ emission is thought to be produced by the heated gas in the magnetospheric accretion flows; therefore, UV excess emission and H$\alpha$ emission probe two distinct, albeit interconnected, components of the magnetospheric accretion process. Several studies have shown a definite correlation between the continuum excess emission of accreting YSO populations and their H$\alpha$ luminosity, measured simultaneously \citep[e.g.,][]{alcala2014,alcala2017}. The H$\alpha$ luminosity is therefore considered a reliable statistical diagnostic of the presence of accretion, although quantitative measurements of the amount of accretion on individual sources can vary depending on the indicator used (e.g., \citealp{antoniucci2011}; \citealp{manara2013a}). As spectroscopic surveys of H$\alpha$ emission can be time-consuming and are often restricted to small and/or nearby YSO populations, photometric surveys in narrow-band H$\alpha$ filters, combined with other optical filters (e.g., $V,r,i$), represent an efficient alternative to provide a statistical map of accretion activity in large cluster populations (e.g., \citealp{demarchi2010}; \citealp{barentsen2011,barentsen2014}; \citealp{biazzo2019}).

\begin{figure}
\centering
\includegraphics[width=\textwidth]{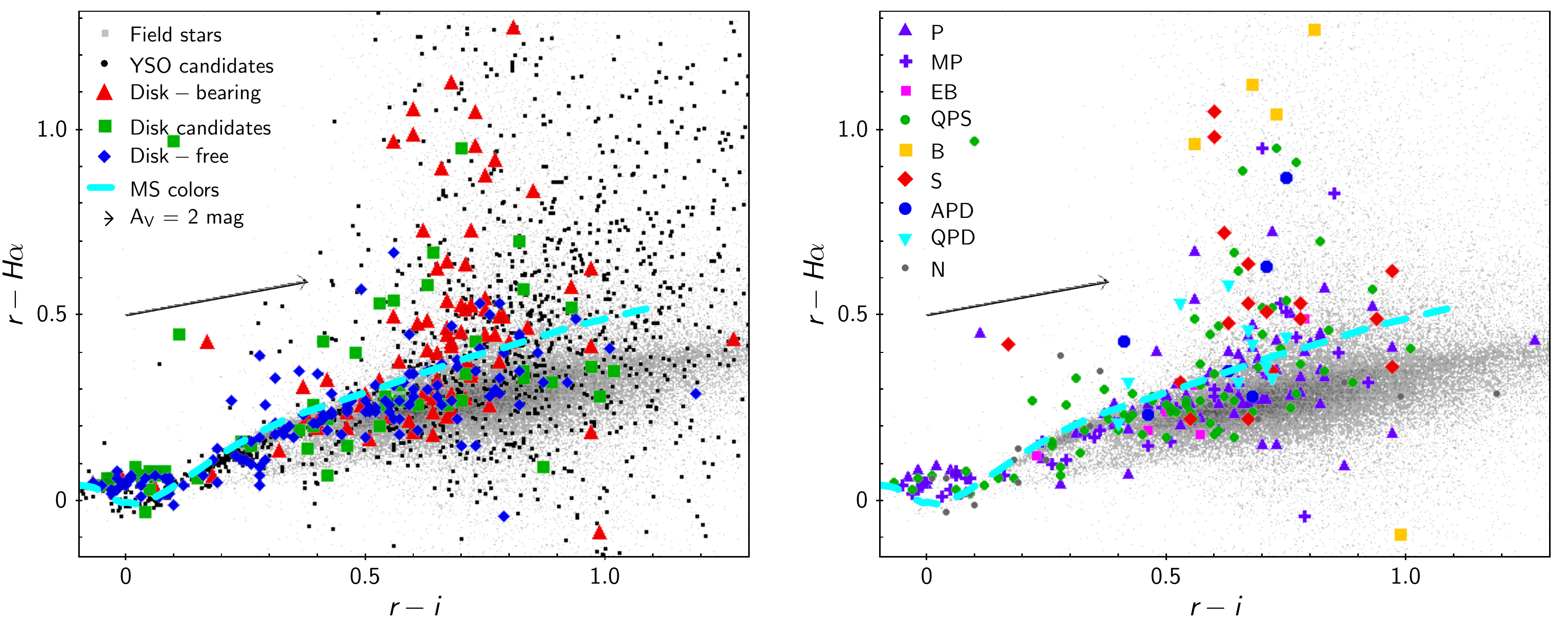}
\caption{{\it Left}: ($r-i$, $r-H\alpha$) color properties of field stars (gray dots) and YSO candidates (black dots) in the Lagoon Nebula region from VST/OmegaCAM photometry. Synthetic colors for main sequence stars in the same filters, as tabulated by \citet{drew2014}, are dashed in cyan to guide the eye. Lagoon Nebula members monitored with {\it K2} and classified as disk--bearing, disk--free, or disk candidates (see Sect.\,\ref{sec:disk_class}) are marked further as red triangles, blue diamonds, and green squares, respectively. The effect of reddening is illustrated as a black vector with $A_V = 2$~mag. {\it Right}: $r,i,H\alpha$ color distribution of {\it K2} Lagoon Nebula members, sorted according to their light curve morphology (see Sect.\,\ref{sec:K2_lc_class}). The color and symbol convention for each class is the same as adopted for Fig.\,\ref{fig:Stetson_J}. Non-variable stars (i.e., those that exhibit flat-line light curves) are shown as large charcoal-gray dots. Disk-bearing stars and irregular {\it K2} variables are shown to exhibit on average higher levels of H$\alpha$ emission (translating to larger $r-H\alpha$ colors) than can be accounted for by photospheric and chromospheric activity.}
\label{fig:ri_rHa}
\end{figure}

Figure~\ref{fig:ri_rHa} illustrates the color properties of field stars and Lagoon Nebula YSOs on the ($r-i$, $r-H\alpha$) diagram. As already observed for the $u$-band properties in Fig.\,\ref{fig:ur_u_disk_K2}, the color loci occupied by disk--bearing and disk--free stars overlap significantly (left panel). However, their distributions exhibit statistical differences, as 56\% of disk--bearing YSOs have $r-H\alpha$ colors above the photospheric level for main sequence dwarfs (i.e., they exhibit an enhanced emission at H$\alpha$ wavelengths), while this percentage is only 27\% among disk--free members and 36\% among disk candidate YSOs\footnote{Objects with $r-i < 0.25$ and $r-H\alpha < 0.2$ were not considered for this computation, since their location above the main sequence track may be caused by non-zero extinction, and de-reddening their photometry would render their colors consistent with the main sequence level.}. Burster stars exhibit the largest $r-H\alpha$ colors (right panel), followed by stochastic stars; in both categories, an excess in H$\alpha$ emission with respect to the photospheric level is detected for $\sim$80\% of the objects. About two thirds of the dipper stars also exhibit redder $r-H\alpha$ colors than the main sequence level. The percentage of YSOs with enhanced $H\alpha$ emission is instead lower among quasi-periodic variables ($\sim$40\%) and strictly periodic sources ($\sim$30\%).

\section{Correlated color-luminosity variability} \label{sec:correlated_lumin_color_var}

Multi-wavelength variability and color monitoring provide key information on the physical drivers of the observed YSO behavior. Indeed, while disk--related phenomena typically induce larger magnitude variations than stable photospheric activity (dark starspots; e.g., \citealp{vrba1993}; \citealp{bouvier1995}; \citealp{grankin2008}; \citealp{venuti2015}), it is the associated color variability that enables discriminating between different physical scenarios. Luminosity modulation by surface features, be them cold magnetic spots or hot accretion shocks, always produce larger variability amplitudes at shorter wavelengths; it is the rate at which the amplitudes decrease towards longer wavelengths that allows us to constrain the physical conditions of the modulating features. Similarly, monitoring the time variability of $u-r$ and $r-H\alpha$ colors (indicative of accretion activity) as the stellar flux evolves provides details on the accretion geometry in disk--dominated YSO variables.

\begin{figure}
\centering
\includegraphics[width=\textwidth]{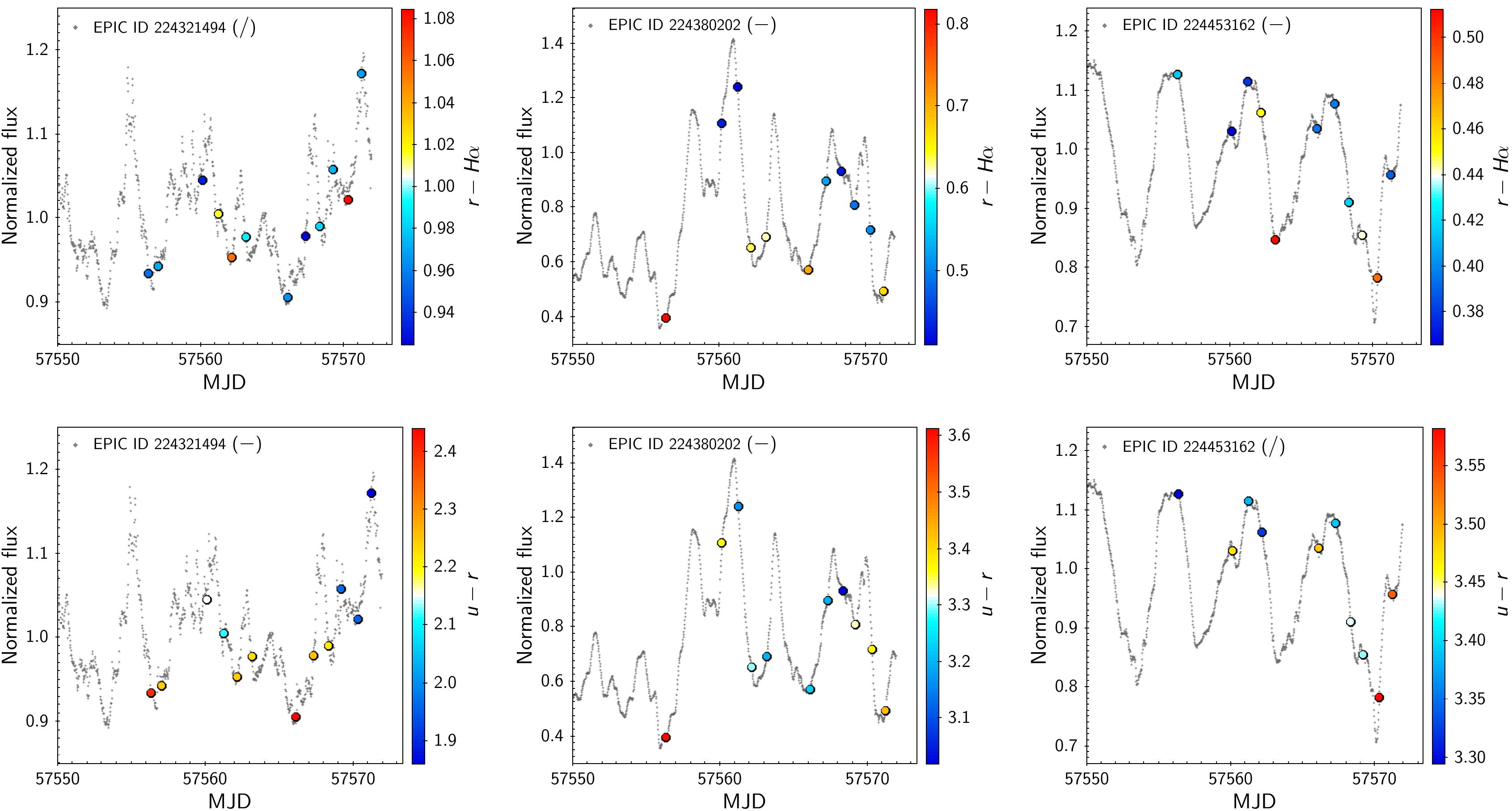}
\caption{Last 25 days of the {\it K2} light curves for selected Lagoon Nebula YSOs, representative of the burster (left panels), stochastic (middle panels), and quasi-periodic dipper (right panels) variable classes, with superimposed measurements of the instantaneous $r-H\alpha$ (top panels) and $u-r$ (bottom panels) colors, obtained simultaneously with VST/OmegaCAM. The range of color values is matched to a color scale, as indicated to the right of each panel. {The state of correlation between flux variations and color variations, tested as described in the text, is indicated on each panel next to the target's EPIC~ID, following the convention adopted in Table~\ref{tab:K2_class_color_accretion} (+ for positive correlation trends, -- for negative correlation trends, and / for non-correlated trends).}}
\label{fig:K2_lc_VST_colors}
\end{figure}

To probe the comparative nature of flux variations in disk--dominated YSOs that belong to distinct {\it K2} morphology classes, we investigated the simultaneous {\it K2} flux and VST/OmegaCAM $u,r,H\alpha$ color variations for Lagoon Nebula objects in our sample that were detected in all filters and exhibit either a burster, a stochastic, or a dipper light curve. As reported in Sects.\,\ref{sec:K2} and \ref{sec:VST}, our VST survey of the Lagoon Nebula overlapped with the last two weeks of the {\it K2} run, and roughly two thirds of the VST observing epochs were acquired during this time window. For each object, we cross-correlated the {\it K2} light curves with the VST color time series, and retained those epochs that match in time within an interval of 0.01~d (i.e., $<$\,15~min), in order to avoid erroneous flux and color associations for short--lived variability phenomena (e.g., bursts with duration of hours; see \citealp{stauffer2014}, \citealp{cody2017}). Examples of how color variations are associated with luminosity variations across our sample are reported in Fig.\,\ref{fig:K2_lc_VST_colors}. 
To investigate any trends between flux variations and color variations, we measured the degree of correlation between simultaneous {\it K2} flux measurements and, in turn, $u-r$ and $r-H\alpha$ colors. We applied a simple least-squares fit model to derive the slope of the best correlation trend, and then evaluated its significance against the average slope and slope dispersion ($\sigma$) obtained for 100 randomly shuffled and recombined sets of the original flux and color measurements for the object. We retained as significant those correlation or anticorrelation trends whose slope stood at least 1~$\sigma$ away from the average slope derived from the 100 simulated sets. The statistical results of this analysis are reported in Table~\ref{tab:K2_class_color_accretion}.

\begin{table}
\caption{Statistics on the fractions of disk-bearing Lagoon Nebula YSOs, with {\it K2} light curves classified as burster (B), stochastic (S), aperiodic dipper (APD), quasi-periodic dipper (QPD), or quasi-periodic symmetric (QPS), that exhibit correlated ($+$), anticorrelated ($-$), or non-correlated ($\cancel{\hspace{10pt}}$) trends between their {\it K2} normalized light curve flux ($f_{norm}$) and the $u-r$ and $r-H\alpha$ color variations monitored with VST/OmegaCam. The number of YSOs, in each light curve category, for which the flux--color trend could be analyzed is reported below each category label in the table header. Rows shaded in gray correspond to an anticorrelation trend between $f_{norm}$ and $u-r$, expected in case of accretion-dominated variability.}
\label{tab:K2_class_color_accretion}
\centering
\setlength{\tabcolsep}{0em}
\begin{tabular}{r l | c c c c c}
\hline
\multicolumn{2}{c|}{\multirow{2}{*}{\mbox{  }$f_{norm}$ trend\mbox{  }}} & {B} & {S} & {APD} & {QPD} & {QPS} \\
\multicolumn{2}{c|}{} & {\small \mbox{    }[4 YSOs]\mbox{  }} & {\small \mbox{  }[15 YSOs]\mbox{  }} & {\small \mbox{  }[6 YSOs]\mbox{  }} & {\small \mbox{  }[10 YSOs]\mbox{  }} & {\small \mbox{  }[25 YSOs]\mbox{    }} \\
\hline
 \multirow{2}{*}{\mbox{  }$\propto$\mbox{  }} & $+(u-r)$ & \multirow{2}{*}{--} & \multirow{2}{*}{--} & \multirow{2}{*}{--} & \multirow{2}{*}{--} & \multirow{2}{*}{--} \\[-1pt]
 & $+(r-H\alpha)$ & & & & & \\
 \hline\hline
 \multirow{2}{*}{\mbox{  }$\propto$\mbox{  }} & $+(u-r)$ & \multirow{2}{*}{--} & \multirow{2}{*}{--} & \multirow{2}{*}{--} & \multirow{2}{*}{--} & \multirow{2}{*}{8.0\%} \\[-1pt]
 & $-(r-H\alpha)$ & & & & & \\
 \hline\hline
 \rowcolor{Gray}
 & $-(u-r)$ &  &  &  & & \\[-1pt]
 \rowcolor{Gray}
 \multirow{-2}{*}{\mbox{  }$\propto$\mbox{  }} & $+(r-H\alpha)$ & \multirow{-2}{*}{--} & \multirow{-2}{*}{20.\%} & \multirow{-2}{*}{--} & \multirow{-2}{*}{--} & \multirow{-2}{*}{4.0\%}\\
 \hline\hline
 \rowcolor{Gray}
 & $-(u-r)$ &  &  &  &  & \\[-1pt]
 \rowcolor{Gray}
 \multirow{-2}{*}{\mbox{  }$\propto$\mbox{  }} & $-(r-H\alpha)$ & \multirow{-2}{*}{75.\%} & \multirow{-2}{*}{6.7\%} & \multirow{-2}{*}{33.3\%} & \multirow{-2}{*}{--} & \multirow{-2}{*}{8.0\%} \\
 \hline\hline
 \multirow{2}{*}{\mbox{  }$\propto$\mbox{  }} & $\cancel{\pm(u-r)}$ & \multirow{2}{*}{--} & \multirow{2}{*}{13.3\%} & \multirow{2}{*}{16.7\%} & \multirow{2}{*}{--} & \multirow{2}{*}{8.0\%} \\[-1pt]
 & $+(r-H\alpha)$ & & & & & \\
 \hline\hline
 \multirow{2}{*}{\mbox{  }$\propto$\mbox{  }} & $\cancel{\pm(u-r)}$ & \multirow{2}{*}{--} & \multirow{2}{*}{6.7\%} & \multirow{2}{*}{--} &   \multirow{2}{*}{40.\%} & \multirow{2}{*}{16.\%} \\[-1pt]
 & $-(r-H\alpha)$ & & & & & \\
 \hline\hline
 \multirow{2}{*}{\mbox{  }$\propto$\mbox{  }} & $+(u-r)$ &\multirow{2}{*}{--}  & \multirow{2}{*}{6.7\%} & \multirow{2}{*}{16.7\%} & \multirow{2}{*}{--} & \multirow{2}{*}{12\%} \\[-1pt]
 & $\cancel{\pm(r-H\alpha)}$ & & & & & \\
 \hline\hline
 \rowcolor{Gray}
 & $-(u-r)$ &  &  &  &  & \\[-1pt]
 \rowcolor{Gray}
 \multirow{-2}{*}{\mbox{  }$\propto$\mbox{  }} & $\cancel{\pm(r-H\alpha)}$ & \multirow{-2}{*}{25.\%} & \multirow{-2}{*}{40.\%} & \multirow{-2}{*}{16.7\%} & \multirow{-2}{*}{30.\%} & \multirow{-2}{*}{20\%} \\
 \hline\hline
 \multirow{2}{*}{\mbox{  }$\propto$\mbox{  }} & $\cancel{\pm(u-r)}$ & \multirow{2}{*}{--} & \multirow{2}{*}{6.7\%} & \multirow{2}{*}{16.7\%} & \multirow{2}{*}{30.\%} & \multirow{2}{*}{24\%} \\[-1pt]
 & $\cancel{\pm(r-H\alpha)}$ & & & & & \\
 \hline
\end{tabular}
\end{table}

A first clear indication from this analysis is that very few objects with disk--dominated variability exhibit a positive correlation between their flux and their $u-r$ colors. This implies that brighter states for these sources typically correspond to enhancements of their $u$-band luminosity (leading to lower, hence bluer, $u-r$ values). An anticorrelation trend between flux and $u-r$ measurements is detected in all burster stars in our sample, and in two thirds of the stochastic stars. This is consistent with our interpretation that the most prominent flux variations exhibited by these sources are associated with discrete accretion events and the resulting shock emission. A definite correlation or anticorrelation trend with $r-H\alpha$ is detected in a smaller number of cases for these two classes, 75\% of our small sample of busters, and around 50\% of our stochastic stars. This is also consistent with the fact that the H$\alpha$ emission provides a more indirect proxy for accretion onto the star than UV emission, and it originates from a more extended region in the inner disk environment. All of the burster stars for which a definite flux vs. $r-H\alpha$ trend is found (75\%) exhibit an anticorrelation between these two quantities, which suggests that the peaks in brightness for these objects correspond to phases of unobstructed view onto the accretion shock regions, when continuum emission is enhanced with respect to the measured line emission (yielding smaller $r-H\alpha$ values). Conversely, one third (33.3\%) of the stochastic stars exhibit a positive correlation between their flux and $r-H\alpha$ values, 2.5 times more numerous than those (13.4\%) which instead exhibit an anticorrelation trend. This suggests a geometry where, at the brightness peak or close to it, both surface accretion features (shocks) and extended accretion funnels are visible. The accretion dynamics in burster stars is believed to be governed by instabilities at the interface between the stellar magnetosphere and the inner disk rim; such instabilities prevent the formation of magnetospheric--driven accretion columns, and instead fuel accretion onto the star via thin, equatorial tongues of material that are only funneled along the magnetic field lines when they are already close to the stellar surface \citep{kulkarni2008}. A somewhat different accretion mechanism was proposed for stochastic stars \citep{stauffer2016}: namely, a variable influx of material feeds into magnetically channeled accretion flows, which translates to stochastically variable mass loads onto the star, and to rapidly evolving hotspot regions. This picture would link stochastic stars to aperiodic dippers seen at lower inclinations, and is consistent with the differing color trends we observe, on a statistical level, between burster stars and stochastic stars.

Among dipper stars, only around half of the aperiodic dippers, and 30\% of the quasi-periodic dippers, exhibit a definite anticorrelation trend between their flux and $u-r$ variations. The decrease in detection of anticorrelated $f_{norm}$ vs. $u-r$ trends from B to QPD variables in Table~\ref{tab:K2_class_color_accretion} may suggest that stochastic stars and aperiodic dippers represent an intermediate mode of star--disk interaction between the instability--driven regime observed in bursters and the stable funnel--flow regime observed in quasi-periodic variables. About a third of aperiodic dippers and the majority of quasi-periodic dippers exhibit no specific trend in $u-r$ as their luminosity varies. This may be a consequence of the fact that, in dipper stars, the most prominent flux variations (recurring luminosity dips) are a product of variable extinction by clumps of material in the inner disk environment, which can effectively mask the luminosity variations induced by accretion shocks. Around 45\% of dipper stars exhibit a definite trend between their flux variations and their $r-H\alpha$ color behaviors; among these, the most common behavior is an anticorrelation trend, which indicates that their $r-H\alpha$ is largest at their lowest luminosity levels (i.e., inside the dips). Larger values of $r-H\alpha$ correspond to an enhancement of the narrow-band H$\alpha$ emission with respect to the $r$-band continuum emission. This trend is qualitatively consistent with a scenario in which flux dips are produced by an inner disk warp at the base of a magnetospheric accretion column, where the phase of minimum photospheric visibility coincides with a maximum in accretion funnel visibility. {The large number of dipper variables for which no specific trends between flux variations and color variations were found may be explained in terms of multiple co-existing accretion streams or columns, and/or assuming a delayed appearance of accretion features and occultation features. The latter may be a common scenario, as suggested by earlier studies \citep{mcginnis2015} where direct evidence of hot spots appearing simultaneously with the main occultation event was recovered only in a minority of cases.}

A redder spectrum at lower luminosity states is also expected for stars with variability dominated by surface spot modulation, as mentioned earlier. Indeed, for spots hotter than the photosphere, the spot spectrum peaks at shorter wavelengths than the stellar spectrum, resulting in an enhanced spot--to--photosphere contrast at bluer wavelengths. Conversely, for spots colder than the photosphere, the spot spectrum peaks at longer wavelengths than the stellar spectrum, resulting again in an enhanced spot--to--photosphere contrast at bluer wavelengths. However, when both cold and hot spots are present at the stellar surface (as is likely for quasi-periodic symmetric variables in our sample), we may expect their photometric signatures to overlap and mask any color correlation. Indeed, as reported in Table~\ref{tab:K2_class_color_accretion}, quasi-periodic symmetric stars are found to exhibit a wide range of color behaviors. Only about a third of YSOs in this category, among our disk-bearing stars, exhibit a clear anticorrelation trend between the {\it K2} flux intensity and the VST $u-r$ color; this percentage is similar to what observed among quasi-periodic dippers. Conversely, about half of the quasi-periodic symmetric variables do not exhibit a definite correlation trend with $u-r$ and/or $r-H\alpha$ colors.



\section{Timescales of variability from {\it K2} light curves} \label{sec:SF}

In order to probe the characteristic timescales of variability of Lagoon Nebula members, irrespective of the periodic or aperiodic nature of their variation patterns, we adopted the method of structure functions (\citealp{simonetti1985}; \citealp{hughes1992}; \citealp{devries2003}), as recently implemented to study YSO variability by \citet{sergison2020}. The method consists of extracting every timescale of variability $\tau$ encompassed by the time series, and, for each $\tau$, computing the average amplitude of normalized flux variability among all pairs of points in the light curve that are spaced by that time interval $\tau$. The structure function ($\mathcal{SF}$) is then defined as the average variability amplitude measured within the light curve as a function of $\tau$. In order to apply the $\mathcal{SF}$ method to a discrete time series, with possibly unevenly spaced data, we first identify the range of investigated $\tau$, from $\tau_{min}$ (i.e., the minimum timescale of variability that can be reliably extracted from the light curve) to $\tau_{max}$ (i.e., the maximum timescale of variability that can be reliably extracted from the light curve). We then divide this ($\tau_{min}$, $\tau_{max}$) range into logarithmically spaced timescale bins, and, for each bin ($\tau_1$, $\tau_2$), we select all pairs of light curve epochs ($t_i$, $t_j$), where $j>i$ and $\tau_1 < t_j - t_i < \tau_2$. The average variation in normalized flux, measured across all pairs of selected points, defines the value of $\mathcal{SF}(\tau_1, \tau_2)$ as
\begin{equation} \label{eqn:SF}
\mathcal{SF}(\tau_1, \tau_2) = \sqrt{\frac{1}{N(\tau_1, \tau_2)} \sum_{j>i}(f_i - f_j)^2}\,,
\end{equation}
where $N(\tau_1, \tau_2)$ is the number of pairs of light curve points ($i,j$) separated in time by an amount comprised between $\tau_1$ and $\tau_2$, and $f_i$ is the normalized flux measured at time $t_i$.

\begin{figure}
\centering
\includegraphics[width=\textwidth]{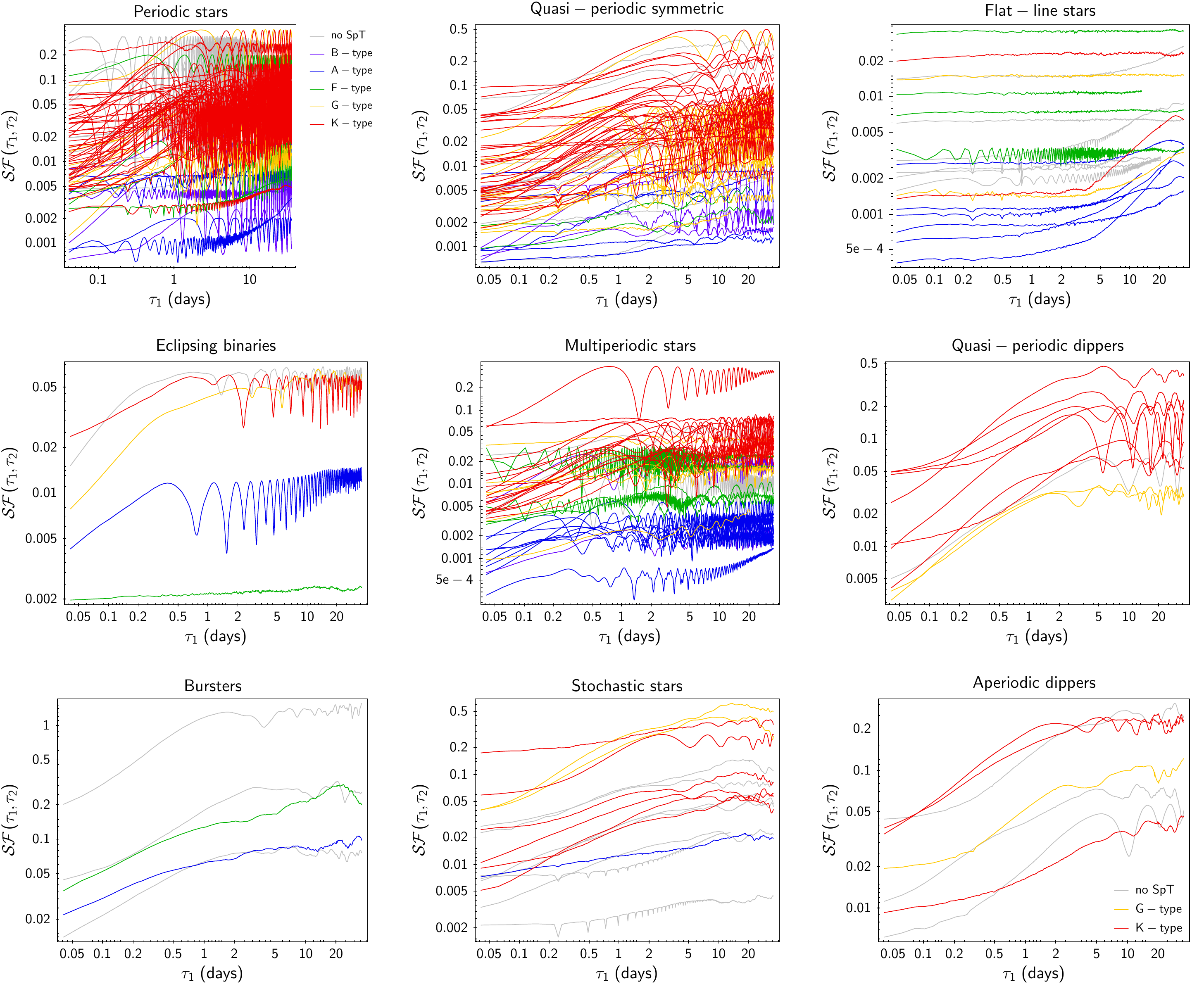}
\caption{Structure functions $\mathcal{SF}$ computed for each {\it K2} light curve on timescales $\tau$ ranging from twice the data cadence to half the light curve span, sampled as discussed in Sect.\,\ref{sec:SF}. For each bin ($\tau_1, \tau_2$) of investigated variability timescales, the value of $\mathcal{SF}$ resulting from Eq.\,\ref{eqn:SF} is plotted as a function of $\tau_1$ for illustration purposes. Each class of light curve variability is shown in a different panel, as indicated on top of each diagram (the `U' class described in Sect.\,\ref{sec:K2_lc_class} is not included here). Different colors identify different SpT, as detailed to the right of the top left panel (purple~$\rightarrow$ B--type stars; blue~$\rightarrow$ A--type stars; green~$\rightarrow$ F--type stars; yellow~$\rightarrow$ G--type stars; red~$\rightarrow$ K--type stars; gray~$\rightarrow$ stars with no SpT estimate from the procedure in Sect.\,\ref{sec:Av_SpT}). On all panels, it is possible to observe a systematic increase of the intrinsic variability amplitude (as traced by the value of $\mathcal{SF}$ in the plateau region at large timescales) from earlier-type stars to later-type stars.}
\label{fig:K2_class_SF}
\end{figure}

From a theoretical standpoint, the behavior of $\mathcal{SF}(\tau)$ is expected to consist of three main separate regimes (see Fig.\,7 of \citealp{sergison2020}). Initially, the $\mathcal{SF}$ is expected to be relatively flat or slowly-increasing with $\tau$, corresponding to the short-$\tau$ regime where the observed flux variations are dominated by photometric uncertainties rather than intrinsic variability. In a second phase, the $\mathcal{SF}$ starts rising above the noise-dominated level and increases as a power law, with a specific gradient that reflects the nature of the observed variability. The power-law rise continues until a limiting value $\tau_{\scriptscriptstyle high}$, which corresponds to the largest timescale at which intrinsic variability is observed along the time series (i.e., the variability observed beyond $\tau_{\scriptscriptstyle high}$ merely reflects the variability displayed on shorter timescales). $\mathcal{SF}(\tau > \tau_{\scriptscriptstyle high})$ therefore exhibits a newly flat or slowly-varying trend around the value of $\mathcal{SF}(\tau_{\scriptscriptstyle high})$. We selected our $\tau_{min}$ to correspond to twice the light curve cadence (i.e., $\tau_{min} \sim 0.04$~d), and our $\tau_{max}$ as half the total light curve span (i.e., $\tau_{max} \sim 35.5$~d). We then sampled this interval in 1000 equal bins in logarithmic space, and for each bin we computed the value of $\mathcal{SF}$ as defined in Eq.\,\ref{eqn:SF}. The resulting $\mathcal{SF}(\tau)$ are illustrated in Fig.\,\ref{fig:K2_class_SF} for different light curve behaviors and spectral classes.

As already discussed in Sect.\,\ref{sec:VST_variability}, later-type stars (shown in yellow and in red on Fig.\,\ref{fig:K2_class_SF}) tend to exhibit systematically higher amounts of variability than earlier-type stars (purple and blue). Flat-line (or non-variable) stars (Fig.\,\ref{fig:K2_class_SF}, top right panel) exhibit either an overall constant $\mathcal{SF}$ throughout the $\tau$ range, or an initially flat trend (up to timescales of a few days) that subsequently increases exponentially with $\tau$ until the end of the timescale domain. This indicates that the flux variations observed for light curves classified as `N' are dominated either by luminosity fluctuations that appear with similar amplitudes throughout the monitored span, or by light curve systematics that impact the observed amplitude of luminosity variations as the considered time span increases. In the other light curve categories, two or three distinct regimes are typically identified in the observed $\mathcal{SF}$, which correspond to the $\tau$ domains dominated respectively by photometric noise, intrinsic variability, and reflected variability from shorter timescales. 

\begin{figure}
\centering
\includegraphics[width=\textwidth]{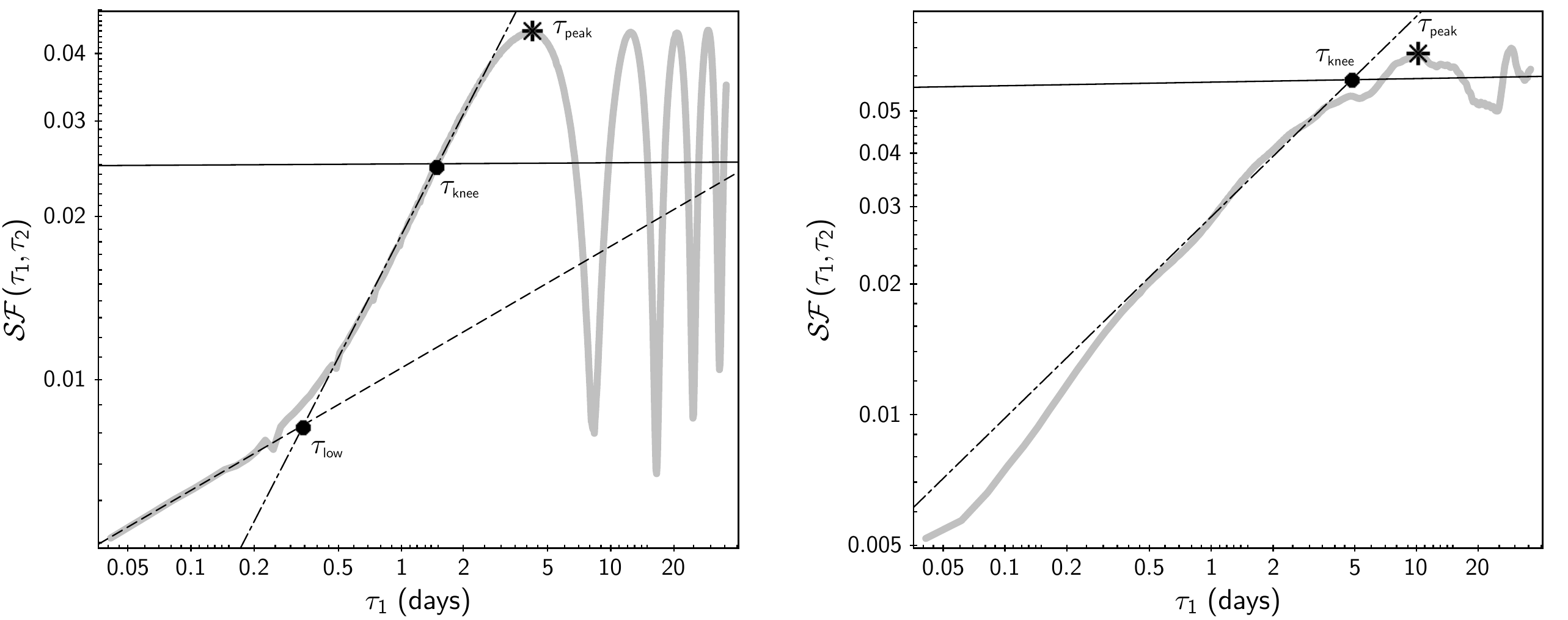}
\caption{{Examples} of $\mathcal{SF}$ analysis conducted on {\it K2} light curves to extract the characteristic timescales of variability. The $\mathcal{SF}$ shown in gray {on the left} belongs to the K--type star EPIC ID 224438171, whose variability behavior is classified as P{, while the one shown on the right belongs to the K--type star EPIC ID 224337699, classified as S}. The dashed black line traces the best power-law fit derived for the first $\mathcal{SF}$ regime ($\tau$ range where the photometric noise is predominant{; not observed in the second case}). The dash-dotted black line traces the best power-law fit derived for the {main} $\mathcal{SF}$ regime ($\tau$ range dominated by intrinsic variability on the corresponding timescales). The solid black line traces the best power-law fit to the third $\mathcal{SF}$ regime ($\tau$ range where the observed variability is a reflection of the variability observed on shorter timescales). The timescales of intersection between first-regime and second-regime fits ($\tau_{\scriptscriptstyle low}$), and between second-regime and third-regime fits ($\tau_{\scriptscriptstyle knee}$) {are marked as large black dots}. The asterisk marks the timescale corresponding to the first observed maximum in the $\mathcal{SF}$ ($\tau_{\scriptscriptstyle peak}$). The small negative spikes{, visible in the $\mathcal{SF}$ on the left} at $\tau \sim 0.24$~d, 0.49~d, and repeatedly after that, are spurious features induced by the corrective thruster firings operated roughly every six hours during the {\it K2} mission \citep{howell2014}.}
\label{fig:SF_analysis_example}
\end{figure}

To investigate the nature and characteristic timescales of variability of each class, we fitted a power law ($\mathcal{SF} \sim k \tau^\beta$) to each distinct segment of $\mathcal{SF}$ for each object, as illustrated in Fig.\,\ref{fig:SF_analysis_example}. We then extracted the coordinates of the intersection points between the separate fits (which correspond to the approximate timescales, $\tau_{\scriptscriptstyle low}$ and $\tau_{\scriptscriptstyle knee}$, where the transition from one $\mathcal{SF}$ regime to the next occurs). {These two timescales of intersection delimit the approximate range of timescales within which intrinsic variability is observed. The oscillations observed beyond $\tau_{\scriptscriptstyle knee}$ reflect the periodicity of the observed variability (a maximum in $\mathcal{SF}$ is measured when the pairs of datapoints separated by the corresponding timescale are collected at opposite variability phases, while a minimum in $\mathcal{SF}$ is measured when the timescale considered is a multiple of the actual variability period in the light curve).} We also extracted the location of the first observed peak in the $\mathcal{SF}$, as well as the slope of the fit to the intrinsic variability--dominated regime, which holds clues to the origin of the observed variability.
A clear three-component fit, with the identification of a $\tau$-region dominated by photometric noise, was obtained in $\sim$60\% of the cases. Many of the cases where no noise--dominated region could be identified in the $\mathcal{SF}$ (i.e., where intrinsic variability is observed already at the shortest investigated $\tau$) correspond to stars where the maximum timescale of intrinsic variability is comparatively short: over 65\% of stars where the first $\mathcal{SF}$ peak is located at $\tau_{\scriptscriptstyle peak} < 1$~d exhibit no noise--dominated regime, while among stars with $\tau_{\scriptscriptstyle peak} \geq 1$~d this percentage is only 16\%. The fraction of stars whose $\mathcal{SF}$ is initially dominated by photometric noise is especially low among multiperiodic stars or eclipsing binaries, possibly due to the coexistence of multiple cyclic variability trends. Another category with no substantial noise-dominated $\tau$-region is that of burster stars, which exhibit the most intense, and often short-lived, flux variations \citep[e.g.,][]{cody2017}. For the other categories of stars, the median $\tau_{\scriptscriptstyle low}$ measured ranges from $\sim$0.2~d to $\sim$0.4~d, and tends to be higher among irregular variables than among regular variables.

The index $\beta$ of the best power-law fit to the $\mathcal{SF}$ region dominated by intrinsic variability can vary broadly from case to case: measured values for our {\it K2} sample of Lagoon Nebula members span the entire range from $<$0.1 to 0.9. As discussed in \citet{sergison2020}\footnote{The definition of the $\mathcal{SF}$ we adopt here (Eq.\,\ref{eqn:SF}) is the square root of the definition adopted by \citet[][Eq.~2]{sergison2020}. Therefore, the reference values for the power-law index $\beta$ discussed in their Table~6 are double the values that need to be considered here for comparison with our fit parameters.}, a $\beta \sim 0$ value is expected in case of a light curve dominated by uncorrelated (white) noise, such as those of stars labeled as `N' in the {\it K2} sample (Sect.\,\ref{sec:K2_lc_class}). Indeed, all of our targets in this class with a power-law fit solution to their $\mathcal{SF}$ have $0 \leq \beta < 0.05$. A power-law index $\beta \simeq 0.05$ describes an $\mathcal{SF}$ dominated by flicker--noise variability \citep{press1978}, characterized by an overall underlying trend of larger variability amplitudes for longer timescales (which can exhibit superimposed, shorter-term coherent or chaotic variability), that has been documented in YSOs (e.g., \citealp{rucinski2008}), and that may be driven, for instance, by fluctuations of the accretion rate in the disk \citep{lyubarskii1997}. Although each class of variables in our sample exhibits a significant internal spread in calculated $\beta$ indices, stars dominated by irregular variability (`S', `B', and `APD') have typical $\beta \sim 0.4-0.5$, above the flicker--noise index and close to the $\beta = 0.5$ index that characterizes Brownian noise (random walk). Higher $\beta$ indices of $\sim 0.6-0.7$ characterize the typical $\mathcal{SF}$ of regular variables (`P', `MP', `QPS', `QPD')\footnote{We exclude EBs from this discussion since the predominant variability signatures on those systems arise not from single--star variability, but from transit events.}. These values stand between the $\beta$ index associated with Brownian noise and that expected for an $\mathcal{SF}$ constructed from a sinusoidal time series ($\beta \sim 1$). 

Stars classified as `B', `QPD', `APD', and `S' exhibit, in this order, the largest typical $\mathcal{SF}$ values measured at $\tau_{\scriptscriptstyle knee}$ across the sample, ranging from ten times to about two times those measured (in decreasing order) for stars classified as `P', `QPS', and `MP'. The median $\tau_{\scriptscriptstyle knee}$ measured across our sample is shortest ($\sim$0.5~d) for stars classified as `P' or `MP', between 1~d and 2~d for (in increasing order) `QPS', `B', and `QPD' stars, and on the order of 3~d for `S' and `APD' stars.
\begin{table}
\caption{Results of a two--sample KS test applied to the cumulative distributions in $\beta$~indices derived from the $\mathcal{SF}$ analysis for stars belonging to different variable classes. $p$-values $\leq$0.05 (corresponding to the threshold where the null hypothesis can be rejected at the 5\%-level) are reported in italic. The value highlighted in boldface correspond to cases where the null hypothesis is also rejected at the 5\%-level according to the Anderson-Darling test statistic. The table is symmetric with respect to its diagonal.}
\label{tab:KS_beta}
\centering
\begin{tabular}{c | l l l l l l l}
\hline
\hline
 & \multicolumn{7}{c}{$p$-values from two--sample KS test} \\
 & \multicolumn{1}{c}{APD} &  \multicolumn{1}{c}{B} &  \multicolumn{1}{c}{MP} &  \multicolumn{1}{c}{P} &  \multicolumn{1}{c}{QPD} &  \multicolumn{1}{c}{QPS} &  \multicolumn{1}{c}{S} \\
 \hline
 APD & & $\cdots$ & $\cdots$ & $\cdots$ & $\cdots$ & $\cdots$ & $\cdots$ \\
 B & 0.7 & & $\cdots$ & $\cdots$ & $\cdots$ & $\cdots$ & $\cdots$ \\
 MP & {\it 0.05} & 0.11 & & $\cdots$ & $\cdots$ & $\cdots$ & $\cdots$ \\
 P & \textbf{\textit{0.008}} & {\it 0.03} & 0.5 & & $\cdots$ & $\cdots$ & $\cdots$ \\
 QPD & 0.18 & 0.3 & 0.9 & 0.6 & & $\cdots$ & $\cdots$ \\
 QPS & 0.06 & 0.14 & 0.9 & 0.09 & 0.8 & & $\cdots$ \\
 S & 0.2 & 0.3 & \textbf{\textit{0.0009}} & \textbf{\textit{0.00007}} & \textbf{\textit{0.008}} & \textbf{\textit{0.0013}} \\
\hline
\end{tabular}
\end{table}
\begin{table}
\caption{Results of a two--sample KS test applied to the cumulative distributions in $\tau_{\scriptscriptstyle knee}$ parameters derived from the $\mathcal{SF}$ analysis for stars belonging to different variable classes. $p$-values $\leq$0.05 (corresponding to the threshold where the null hypothesis can be rejected at the 5\%-level) are reported in italic. The value highlighted in boldface correspond to cases where the null hypothesis is also rejected at the 5\%-level according to the Anderson-Darling test statistic. The table is symmetric with respect to its diagonal.}
\label{tab:KS_tau}
\centering
\begin{tabular}{c | l l l l l l l}
\hline
\hline
 & \multicolumn{7}{c}{$p$-values from two--sample KS test} \\
 & \multicolumn{1}{c}{APD} &  \multicolumn{1}{c}{B} &  \multicolumn{1}{c}{MP} &  \multicolumn{1}{c}{P} &  \multicolumn{1}{c}{QPD} &  \multicolumn{1}{c}{QPS} &  \multicolumn{1}{c}{S} \\
 \hline
 APD & & $\cdots$ & $\cdots$ & $\cdots$ & $\cdots$ & $\cdots$ & $\cdots$ \\
 B & 0.5 & & $\cdots$ & $\cdots$ & $\cdots$ & $\cdots$ & $\cdots$ \\
 MP & \textbf{\textit{0.00013}} & \textbf{\textit{0.013}} & & $\cdots$ & $\cdots$ & $\cdots$ & $\cdots$ \\
 P & \textbf{\textit{0.0012}} & \textbf{\textit{0.05}} & 0.6 & & $\cdots$ & $\cdots$ & $\cdots$ \\
 QPD & 0.2 & 0.6 & \textbf{\textit{0.00011}} & \textbf{\textit{0.0002}} & & $\cdots$ & $\cdots$ \\
 QPS & \textbf{\textit{0.03}} & 0.8 & \textbf{\textit{0.000016}} & \textbf{\textit{0.0003}} & 0.16 & & $\cdots$ \\
 S & 0.8 & 0.4 & \textbf{\textit{0.00000012}} & \textbf{\textit{0.000002}} & 0.10 & \textbf{\textit{0.018}} \\
\hline
\end{tabular}
\end{table}
To test how similar or dissimilar different variable classes are with respect to the derived $\mathcal{SF}$ parameters, we applied a two--sample Kolmogorov--Smirnov (KS) test \citep{press1992} and a two-sample Anderson-Darling test \citep{pettitt1976} to the distributions in $\beta$ and $\tau_{\scriptscriptstyle knee}$ obtained for each {\it K2} class. Results of this analysis are reported in Tables~\ref{tab:KS_beta} ($\beta$ index) and \ref{tab:KS_tau} ($\tau_{\scriptscriptstyle knee}$ parameter). While the significance that can be achieved when comparing two given classes is somewhat dependent on how populous those classes are (and this varies considerably from one class to another, as can be deduced from Table~\ref{tab:K2_class_disks}), some interesting indications can be extracted from these statistical tests. The $\tau_{\scriptscriptstyle knee}$ parameter appears to be the one that discriminates most among different types of variability. A clear distinction is observed between accretion disk--related variability patterns (bursters, stochastic, dippers) and patterns dominated by spot modulation: in these cases, the $p$-values from the KS test indicate that the null hypothesis can be rejected to a significance $\leq$0.05 and often below 1\%\footnote{The higher $p$-values obtained when the `B' class is involved, compared to those obtained when comparing other classes of irregular vs. regular variables, are likely affected by the very small number of stars in our sample with a bursting behavior.}. The same results are obtained when using the Anderson-Darling test. The null hypothesis is instead retained for the comparative distributions in $\tau_{\scriptscriptstyle knee}$ exhibited by `QPS' and `QPD' variables; this suggests that the physical origins for these two categories are at least partly connected. The comparison between different light curve classes in terms of the $\beta$ parameter is more inconclusive, but it suggests a marked difference at least between the most irregular variables and the most regular ones. 

\section{Discussion} \label{sec:discussion}

\subsection{Mass dependence of variability in young stars} \label{sec:var_mass_dependence}

As illustrated in Sects.~\ref{sec:Jindex_spt} and \ref{sec:SF}, the amount of variability exhibited by YSOs in our sample depends strongly on stellar mass or spectral type. This is observed not only within individual categories of light curves with discernible variability patterns, but also among light curves classified as non-variable or flat-line: indeed, as shown in Fig.\,\ref{fig:K2_class_SF} (top-right panel), the baseline of photometric fluctuations measured for B and A stars with light curves classified as `N' is systematically lower than that measured for the light curves of flat-line F-to-K stars. A flat light curve might in principle be associated with YSOs seen in a geometric configuration close to pole-on (i.e., line of sight to the star nearly aligned with the stellar rotation axis), where the side of the star that faces the observer remains essentially the same at different rotational phases. However, if flat-line light curves were driven purely by the geometric viewing angle, we would expect similar fractions of stars classified as `N' to appear at all spectral types, and this is not supported by the data in Table~\ref{tab:K2_class_SpT}. In fact, the fraction of `N' stars is systematically higher among earlier-type\footnote{Data reported in Table~\ref{tab:K2_class_SpT} might seem to suggest a possible discrepant behavior for B-type stars, with a large fraction of detected periodic behaviors, and a very low statistical fraction of `N' stars. However, this may be driven by the limited number of B-type stars in our sample, and the light curve classification for objects belonging to this spectral range is somewhat more uncertain than for other spectral types, as evidenced by the large fraction of sources labelled `U' (unclassifiable).} stars (down to F spectral types) than among later-type (G, K) stars, and this is accompanied by an increase in the fraction of objects with detected modulated behaviors (`P', `QPS') from early-type to late-type stars. These results can be interpreted in terms of distinct magnetic properties for young stars of different masses. As mentioned in Sect.\,\ref{sec:intro}, homogeneous spectropolarimetric surveys of intermediate-mass YSOs \citep{villebrun2019} have revealed very few detections of magnetic fields for objects with effective surface temperatures hotter than $\sim$6\,000~K (which correspond to SpT~$\lesssim$ mid-to-late F; \citealp{herczeg2014}). When magnetic fields are detected, the complexity of the magnetic field structure appears to increase across the H-R diagram, from cooler and purely convective stars that exhibit largely dipolar and axisymmetric fields, to hotter stars with large radiative cores that exhibit more chaotic field configurations with weak dipolar components \citep{gregory2012,hill2019}. As dark starspots are formed at the base of strong stellar magnetic loops that suppress local convective flux motions, we would naturally expect to observe more intense modulated variability when large, localized spots, associated with ordered dipolar fields, are present. Conversely, in a chaotic field configuration where magnetic loops and associated starspots are distributed homogeneously across the stellar surface, the starspot-to-photosphere contrast is expected to be similar at all rotational phases, therefore yielding little modulated variability, irrespective of the stellar inclination \citep[e.g.,][]{kesseli2016}. Our results are consistent with the findings by earlier studies of modulated variability in PMS stars, which noted a marked decrease in the occurrence of spot-induced flux variations from fully convective stars to stars with extended radiative cores within individual populations \citep[e.g.,][]{saunders2009}.

Table~\ref{tab:K2_class_SpT} also reveals a clear distinction between higher-mass and lower-mass stars with respect to disk-driven variability behaviors. The dependence on spectral class is apparent in particular among burster stars (dominated by intense and short-lived accretion), clustered at A and F spectral types, and dipper stars (dominated by occulting dust in the circumstellar environment), clustered at G and K spectral types. A closer inspection of the $\mathcal{SF}$ trends reconstructed for all light curves labelled as `B' (bursting; see Fig.\,\ref{fig:K2_class_SF}, bottom-left panel) suggests that the SpT-dependence observed for this category of variables may be caused, at least partly, by the small number of objects that match this class of photometric behaviors in our sample, and by the fraction of those without an SpT estimate from our analysis in Sect.\,\ref{sec:Av_SpT}. Indeed, as shown in Fig.\,\ref{fig:K2_class_SF}, all stars with a burster-like {\it K2} light curve and with no SpT estimate (marked in gray) exhibit a level of $\mathcal{SF}$ variability that is comparable to, or higher than, what is measured for burster stars with available spectral classification. This feature, combined with the mass-dependent trends in the amount of YSO variability discussed earlier, would place the sources in gray at similar or later spectral classes than the sources with available SpT. Dipper stars with available SpT, on the other hand, delimit the range of $\mathcal{SF}$ properties within which all dipper stars with no spectral classification are also located (Fig.\,\ref{fig:K2_class_SF}, middle-right and bottom-right panels). This suggests that the SpT dependence in dipping behaviors reported in Table~\ref{tab:K2_class_SpT} is real, and the dipping phenomenon is therefore observed only among later-type stars. A trend analogous to the one observed here had already been noted in earlier studies on dipper YSO variables, albeit targeting less massive regions \citep{ansdell2016a}, and a detailed discussion on its possible origin has been presented by \citet{bodman2017}. Namely, if the dipping behavior is produced by inner disk warps or dust entrained in accretion streams that corotate with the star (as suggested by the similarity in periodicity measured for dipping events and modulated variability among `QPD' variables; e.g., \citealp{stauffer2015,mcginnis2015}), the phenomenon is only possible if the disk temperature at the corotation radius is below the dust sublimation temperature, $\sim$1300--1400~K \citep{kobayashi2011,nagel2013}. Assuming that the disk temperature close to the inner edge is driven primarily by direct irradiation from the star, the temperature of a disk annulus in Keplerian rotation around the star depends on the stellar size and effective temperature $T_{\rm eff}$, and on the radial distance from the star, which, in turn, is related to the local orbital velocity (hence orbital period) of the disk material. Under these assumptions, we can then estimate the minimum period at which it would be possible to observe photometric signatures of occulting dust.

\begin{figure}
\centering
\includegraphics[width=0.5\textwidth]{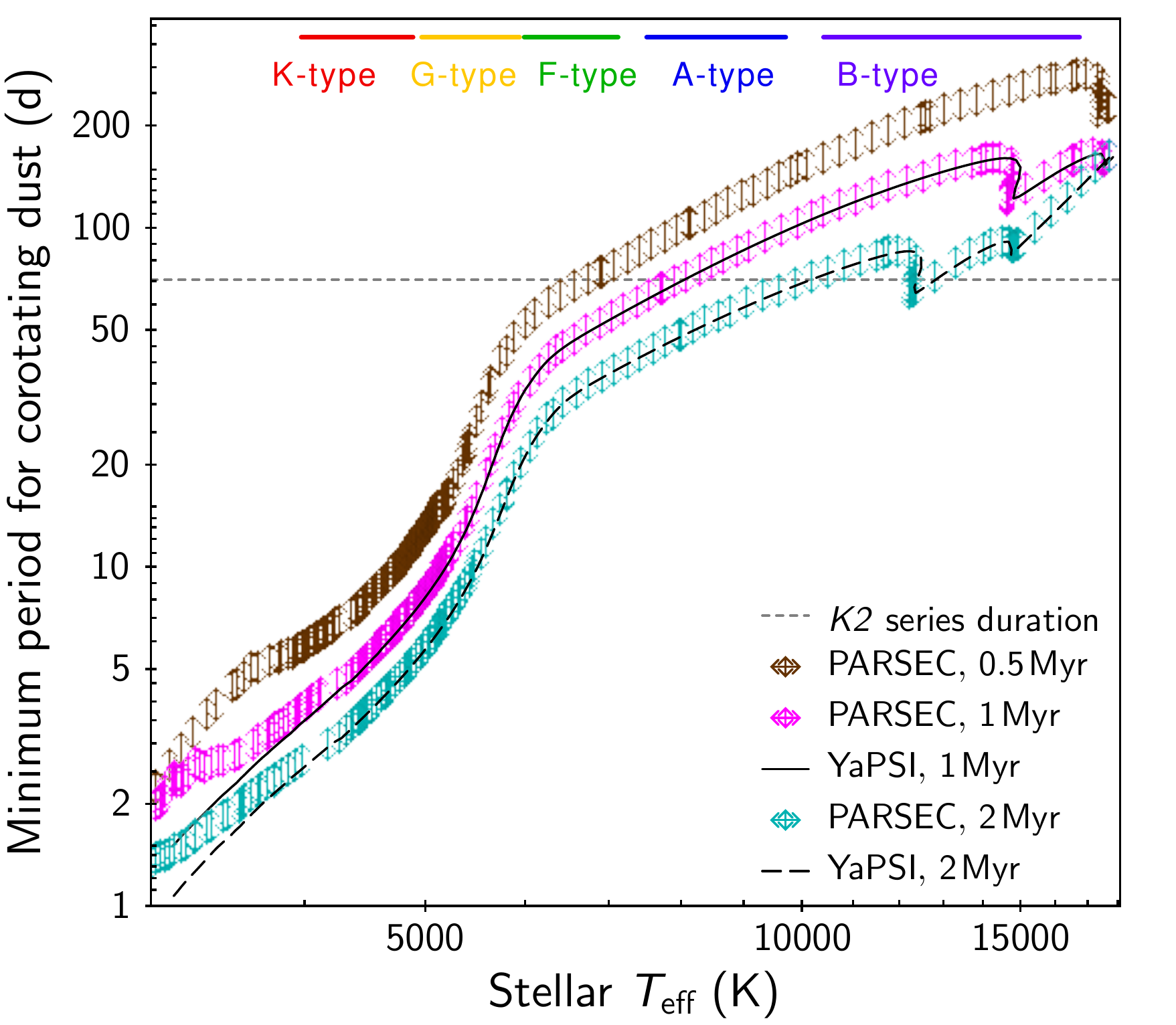}
\caption{Estimates of the orbital period corresponding to the minimum radial distance from the star at which corotating dust can be present in non-sublimated form, as a function of the temperature of the central star. The double-arrows delimit the range of periods calculated, for a given $T_{\rm eff}$, assuming a dust sublimation temperature of 1300--1400~K, and using stellar parameters (mass, luminosity) tabulated by \citet[][PARSEC isochrones]{marigo2017} for 0.5~Myr (brown), 1~Myr (magenta), and 2~Myr (cyan) ages. The average $T_{\rm eff}$--period estimates derived by using \citeauthor{spada2017}'s (\citeyear[][YaPSI]{spada2017}) model isochrones at 1~Myr (solid line) and 2~Myr (dashed line) are also shown to illustrate the overall agreement between different PMS models as a function of stellar $T_{\rm eff}$. The total duration of the {\it K2} time series is shown for comparison purposes as a dotted gray line. Horizontal bars on top of the diagram illustrate the approximate $T_{\rm eff}$ ranges for stars in our sample belonging to different spectral classes.}
\label{fig:Teff_Pcor}
\end{figure}

In Fig.\,\ref{fig:Teff_Pcor}, we illustrate the $T_{\rm eff}$-dependent minimum periods on which a dipping behavior can be observed, calculated from Eqs.\,1 and 2 of \citet{bodman2017}, using 0.5--to--2~Myr theoretical isochrones from the Padova and Trieste Stellar Evolutionary Code (PARSEC) models \citep{marigo2017} and the Yale-Potsdam Stellar Isochrones (YaPSI; \citealp{spada2017}). This diagram reproduces and expands Fig.\,2 of \citet{bodman2017} towards higher $T_{\rm eff}$ values, in order to encompass the entire range of stellar properties relevant to our YSO sample. At any given $T_{\rm eff}$, a dipping behavior can only take place on timescales above the threshold curve traced for a certain age. Periods shorter than (i.e., below) the curve correspond to orbits that are located too close to the star, where the local temperature is too high to allow dust in non-sublimated form. As the diagram illustrates, for lower-mass stars, the minimum periods required for dust at corotation are on the order of a few days, quite consistent with the rotation rates measured for young, disk-bearing stars \citep[e.g.,][]{venuti2017,roquette2017,rebull2020}, which indicates that dipping behaviors are indeed to be expected over weeks-to-months-long monitoring campaigns. As we move to higher $T_{\rm eff}$, however, the period corresponding to the minimum distance from the star where dust can exist increases rapidly, reaching orders of several tens of days by the G-type to F-type transition. These timescales lie at the upper end of the range of rotational periods measured among young stellar populations, which suggests that the actual occurrence of dust structures at the corotation radius around more massive stars ought to be rare. As $T_{\rm eff} \sim 10\,000$~K are reached, in the A--to--B spectral class regime, the minimum period for dust at the corotation radius becomes $\sim 100$~d for $\sim$1~Myr-old YSOs, well beyond the typical rotation rates measured for young stars, and beyond the intrinsic timescales that can be probed with our {\it K2} time series. 

An additional factor that likely plays a role in the disappearance of dipping behaviors at SpT\,$\lesssim$\,G is the increasingly complex magnetic field structure for YSOs at the fully convective to partly radiative transition, as discussed earlier. Indeed, as the star-disk interaction is dominated by the dipole component of the magnetic field (which is the component that exhibits the slowest decline in intensity with radial distance from the star; \citealp{gregory2012}), a strong dipolar field is crucial to establish a stable accretion regime, which in turn leads to longer-lived accretion structures (inner disk warps, accretion columns) responsible for the repeated occultation events (flux dips) observed on high-inclination star-disk systems (\citealp{mcginnis2015}, and references therein). As more massive stars ($\gtrsim 2 M_\odot$) rapidly develop radiative cores, the magnetic pressure on the inner disk regions decreases and the disk is able to push closer to the star, enhancing the spin-up torque on the central object \citep{gregory2012}. This scenario hampers further the possibility of ordered dust structures surviving at the corotation radius around early-type stars, as evident from Fig.\,\ref{fig:Teff_Pcor}. 

\subsection{Geometries of star-disk interaction in young stars}


The variety of photometric behaviors identified among disk-bearing YSOs is expected to reflect a range of star-disk interaction phenomena, and coordinated analyses of the color properties characteristic of different behaviors are critical to assess where and how the observed flux variations are triggered in the star-disk environment. In particular, the IR properties observed for Lagoon Nebula YSOs and depicted in Fig.\,\ref{fig:NIR_MIR_colors_K2} suggest that bursting, stochastic, and dipping behaviors do not share the same origin, although all being driven by disk processes rather than photospheric features. 

Burster stars stand out with respect to other disk-bearing variables in their UV (Fig.\,\ref{fig:ur_u_disk_K2}, right) and H$\alpha$ (Fig.\,\ref{fig:ri_rHa}, right) excess properties, and they also exhibit the largest IR excesses up to wavelengths $\sim$4.5~$\mu$m (Fig.\,\ref{fig:NIR_MIR_colors_K2}). However, the IR properties at wavelengths longer than 4.5~$\mu$m appear consistent with the typical color locus for Class~II sources, and the characteristic timescales of variability extracted from the {\it K2} light curves for bursters are overall consistent in magnitude with those measured for QPS and QPD variables (Sect.\,\ref{sec:SF}). The IR color trends we observe here for burster stars are supported by those inferred in the study of \citet{findeisen2013}, who used data from {\it Spitzer}'s instruments IRAC (3.6--8.0~$\mu$m) and MIPS (24~$\mu$m) to derive a statistical correlation between IR excess and bursting behaviors at shorter wavelengths but not at longer wavelengths. Taken together, these properties indicate that bursting behaviors are driven by the inner disk. Indeed, magnetohydrodynamic simulations of burst-like variability in YSOs have shown that bursting events are triggered by equatorial accretion tongues that evolve rapidly on the inner disk dynamical timescale \citep{kulkarni2008,kulkarni2009}.

Stochastic stars also exhibit a clear association with disk-related activities, notably mass accretion, as indicated by the large fraction of YSOs in this category with enhanced H$\alpha$ emission (Sect.\,\ref{sec:lc_Ha_emission}). However, as shown in Fig.\,\ref{fig:ur_u_disk_K2} (see also \citealp{stauffer2016}), they do not exhibit the strong UV excess levels typical of burster stars, which suggests the erratic behavior of the former may not be driven by the intense accretion regime characteristic of the latter, triggered by instabilities at the inner disk boundary. As mentioned in Sect.\,\ref{sec:lc_IR_colors}, the colors illustrated in Fig.\,\ref{fig:NIR_MIR_colors_K2} hint at distinctive IR emission properties for stochastic stars, which exhibit the reddest shift compared to other disk-bearing YSOs (including bursters) at the longest wavelengths shown on the diagram (IRAC 5.8--8.0~$\mu$m). Similar trends were noted among disk-bearing YSO variables in $\rho$~Ophiucus and Upper~Scorpius by \citet{cody2018}, who employed IR data from $\sim$1.2~$\mu$m to 22~$\mu$m, and reported a segregation in color between erratic light curve variables and other (dipper, modulated) light curve variables at the longest IR wavelengths, but not at near-IR wavelengths, albeit without a clear separation between buster stars and stochastic stars. Similarly, \citet{stauffer2016} reported a comparable color distribution in IRAC filters for bursters and stochastic stars in the NGC~2264 cluster, with a weak indication of stronger dust contribution for stochastic stars than for burster stars, as indicated by the measured slope of their spectral energy distributions between 2~$\mu$m and 8~$\mu$m. Strikingly, stochastic stars also appear to exhibit the longest intrinsic timescales of variability from our analysis in Sect.\,\ref{sec:SF}, therefore suggesting that the origin of their photometric behavior may lie in disk regions beyond the inner rim. 

To explain the variability observed for stochastic YSOs, \citet{stauffer2016} hypothesized that the irregular fluctuations trace a time-variable efficiency in funneling mass inwards from the disk regions around the outer magnetosphere, beyond the co-rotation radius, where the gradient in rotational velocity between disk and magnetosphere triggers the ejection of disk material from the system (propeller regime). A similar mechanism had previously been proposed to explain the accretion variability measured on AA~Tau, the prototype for dipper-like YSO variables \citep{donati2010}, and it had been described theoretically by \citet{romanova2005}. In this scenario, the disk can undergo cyclic variations, being pushed closer to the star and then further out, following the interaction with the rapidly rotating magnetosphere. A mass exchange occurs between the disk and the stellar magnetospheric region along this cycle, with part of the material being expelled via outflows, and a fraction being accreted onto the star. The amount of material successfully channeled towards the star can exhibit rapid variations, with characteristic timescales several times larger than the dynamical timescales of the innermost disk regions ($\sim$few stellar radii from the star), thus potentially explaining the longer dominant timescales of variability measured for stochastic stars in our sample compared to other types of disk-bearing YSO variables. Interestingly, aperiodic dipper variables in our sample are also found to exhibit dominant timescales of variability of the same order as stochastic stars, and longer than those measured for quasi-periodic dippers, burster stars, and quasi-periodic symmetric variables. This suggests that similar mechanisms of variable accretion, coupled with distinct geometric viewing angles (lower inclinations vs. higher inclinations) may drive stochastic and aperiodic dipping behaviors, while their regular counterparts (QPS and QPD) may be driven by stable accretion, with a steady flow of matter from the inner disk onto the star, and a variability pattern modulated by hot spots or accretion structures on the inner disk timescales \citep[e.g.,][]{romanova2004}.

\subsection{Stability of the different modes of YSO variability}

As reported in Table~\ref{tab:K2_class_disks}, 80\% of Class~III objects, and nearly 70\% of Class~II objects in our sample, exhibit some degree of periodicity in their light curves. This indicates that the underlying physical causes for their variability are intrinsically stable at least over timescales of months, albeit with smaller-scale irregular variations that can often be observed on top of the repeated variability profiles for disk-bearing stars, and that likely trace the structure of individual accretion streams. This is consistent with results from the structure function analysis reported in Sect.\,\ref{sec:SF}, from which dominant timescales of variability on the order of days were extracted across our sample. These findings are mirrored by earlier, long-term monitoring studies of variability in young stars \citep[e.g.,][]{grankin2007,grankin2008}, which have revealed very stable photometric patterns for disk-free YSOs over multiple years (corresponding to thousands of rotation periods), with subtle variations in the surface spot properties that are reminiscent of the magnetic cycles observed on our Sun \citep{grankin2008}. Those studies have also revealed a relatively stable typical behavior of young, disk-bearing stars on similar timescales \citep[see also][]{costigan2014,venuti2015}. \citet{venuti2014} examined the comparative extent of variations in accretion luminosity exhibited by a representative sample of young stars in the NGC~2264 cluster on hour-to-weeks timescales. They concluded that such variations are dominated by accretion shock modulation along the stellar rotation period, although changes in the accretion rate throughout the disk and close to the star \citep[e.g.,][]{sergison2020}, as well as dust clumps and density perturbations within the disk \citep[e.g.,][]{bouvier2013}, may induce additional variability beyond the rotational timescales for at least a fraction of Class~II YSOs.


\begin{figure}
\centering
\includegraphics[width=\textwidth]{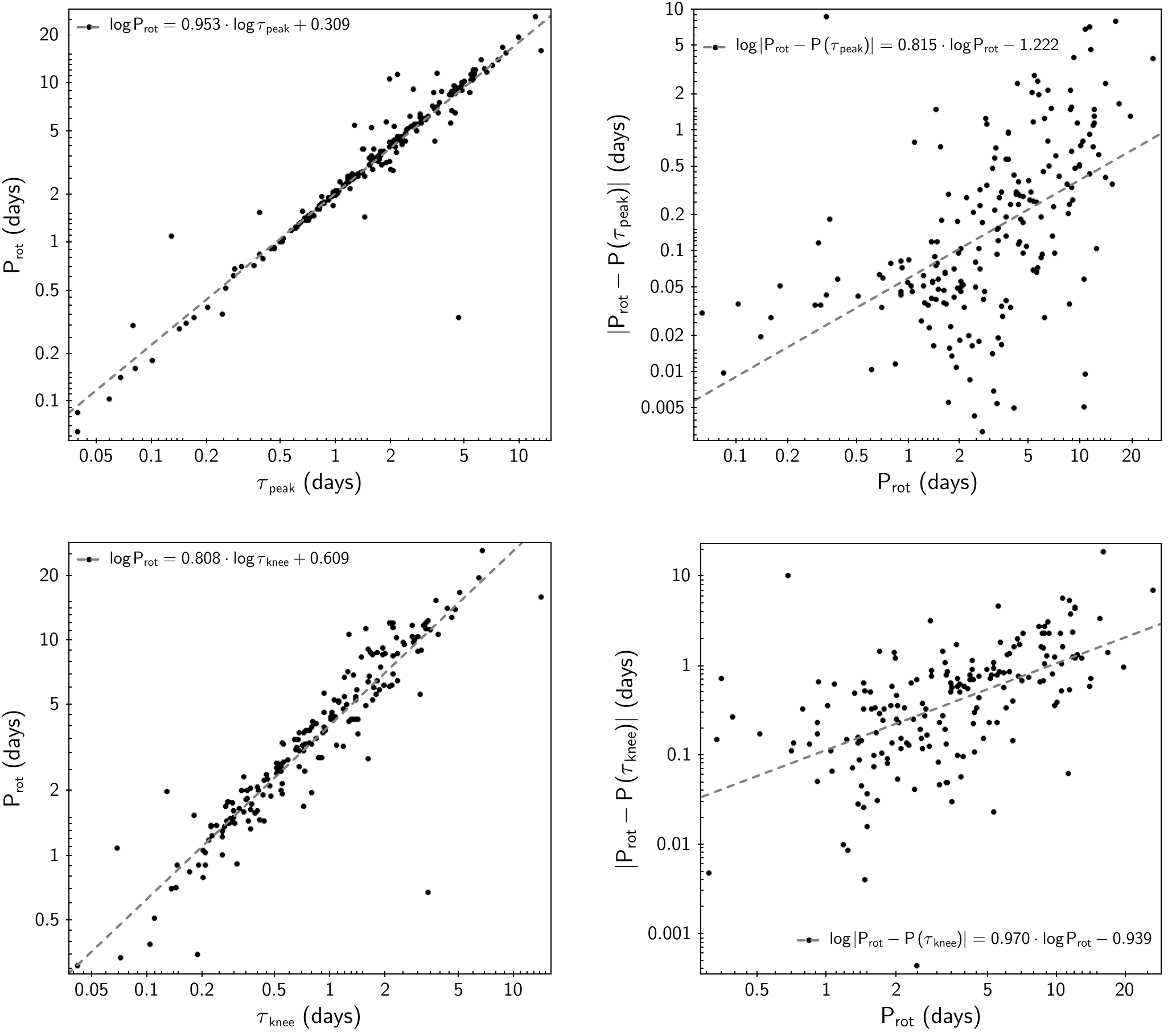}
\caption{{\it Left}: Correlation trends between the rotation period $P_{rot}$ measured from the light curves (Rebull et al., in prep.) and the $\tau_{\scriptscriptstyle peak}$ ({\it top}) and $\tau_{\scriptscriptstyle knee}$ ({\it bottom}) parameters extracted from $\mathcal{SF}$ analysis (see Fig.\,\ref{fig:SF_analysis_example}). The dashed gray lines trace the logarithmic least-squares fit to the datapoint distributions, as indicated in the top-left corner of each diagram. {\it Right}: Absolute residuals between the measured $P_{rot}$ and the period values estimated from $\tau_{\scriptscriptstyle peak}$ ({\it top}) and $\tau_{\scriptscriptstyle knee}$ ({\it bottom}), plotted as a function of $P_{rot}$. The dashed gray lines mark the logarithmic least-squares fit to the distributions of residuals, as labeled on each diagram.}
\label{fig:Prot_taupeak_tauknee}
\end{figure}

To assess the relationship between the characteristic timescales of variability determined for our Lagoon Nebula sample and the actual rotation properties of individual stars, we compared the values of $\tau_{peak}$ (that corresponds to the first observed maximum in the $\mathcal{SF}$, beyond which the $\mathcal{SF}$ typically exhibits a rather flat and oscillatory behavior) with the rotation period of the stars ($P_{rot}$), measured independently from the {\it K2} light curves as part of a companion paper (Rebull et al., in preparation). This comparison, shown in Fig.\,\ref{fig:Prot_taupeak_tauknee} (upper panels), reveals a tight correlation between the two quantities. Similarly, a comparison between the estimated $\tau_{\scriptscriptstyle knee}$ from $\mathcal{SF}$ power-law fitting (i.e., the intersection between the $\mathcal{SF}$ regimes dominated by direct variability and reflected variability, as detailed in Sect.\,\ref{sec:SF}) and the measured $P_{rot}$ by Rebull et al. (in prep.) exhibits a direct correlation between the two quantities across our sample, as shown in Fig.\,\ref{fig:Prot_taupeak_tauknee} (lower panels). The typical $P_{rot}$--to--$\tau_{\scriptscriptstyle knee}$ ratio is $4.5\pm1.6$, consistent with what observed by \citet{sergison2020} for their sample in the Cep~OB3b association.


 The photometric measurement of the rotation period of a star from its light curve (via, e.g., the Lomb-Scargle periodogram technique; \citealp{scargle1982}) relies on the presence of a repeated luminosity modulation pattern, with a discernible periodicity that can be associated with the rotation motion of the star around its own axis. As discussed in Sect.\,\ref{sec:K2_lc_class}, irregular variables in our sample (`B', `S', `APD') generally do not exhibit any clear periodicity in their light curves, albeit characterized by repeated features. Hence, these categories of stars are not represented in the subsample shown in Fig.\,\ref{fig:Prot_taupeak_tauknee}, from which the connection between $\tau_{\scriptscriptstyle knee}$ and $P_{rot}$ is established. If the same connection can be assumed to be true for all stars, the observed trend in typical $\tau_{\scriptscriptstyle knee}$ values suggests that stars whose dynamics are dominated by star--disk interaction phenomena (notably `B', `S', `QPD', `APD') tend to rotate more slowly than stars dominated by geometric modulation by uneven surface features (spots; `P', `QPS', `MP'). This is consistent with the distinct distributions of rotation rates that have been reconstructed for several disked, as opposed to non-disked, young stellar populations \citep[e.g.,][]{venuti2017,roquette2017,rebull2018}, and it can be understood in the presence of a disk-locking mechanism that prevents the star from spinning up until the inner disk has evolved \citep[e.g.,][]{herbst2005,vasconcelos2015,landin2016}. In any case, our analysis indicates that: 
 \begin{enumerate}[label=\roman*)]
 \item for both disk-bearing and disk-free young stars, the most prominent timescale of variability ($\tau_{\scriptscriptstyle knee}$) is on the order of days, which corresponds to the rotational timescales; 
 \item  disk--related phenomena exhibit longer variability cycles than the rotation--induced modulated variability observed on YSOs with little, if any, disk--related activity;
 \item for every type of variable stars, additional variability beyond $\sim \tau_{\scriptscriptstyle knee}$ timescales and up to timescales of months is typically limited (about some percents of the normalized variability observed up to $\tau_{\scriptscriptstyle knee}$). 
\end{enumerate}

These conclusions are at least qualitatively consistent with the results obtained by \citet{sergison2020} in their exploration of a wider range of YSO variability timescales, extending to several years. Of note, those authors reported that about one third of Class~II objects in their sample do exhibit additional variability beyond the days-to-weeks timescales. A discrepancy between the maximum timescale of variability extracted from their structure function analysis and the rotational properties measured for stars in their sample emerges at $\tau$ longer than $\sim10$~d (see their Fig.\,14), which corresponds to the upper end of the $\tau_{knee}$ range derived here. The Class~II objects with long-term variability in \citeauthor{sergison2020}'s (\citeyear{sergison2020}) sample were found to exhibit smaller amplitudes of variability than Class~II objects with no additional variability beyond the rotational timescales. \citet{sergison2020} concluded that these objects likely exhibit limited modulated variability, due to either complex magnetic field geometries or low-inclination viewing angles to the stars. This reduction in the typically predominant modulated variability would allow additional sources of variability to emerge on longer timescales. Interestingly, no $\tau_{knee} \gtrsim 10$~d were detected in our sample, in spite of our ability to detect intrinsic $\tau \lesssim 35$~d from the {\it K2} light curves (see Sect.\,\ref{sec:SF}). This suggests that any additional variability driven by perturbations or accretion across the disk would manifest on timescales longer than that amount (corresponding to several rotational cycles for these young stars). This, in turn, provides some constraints on the minimum duration of the stability cycles intrinsic to these processes, and/or on the location within the disk where longer-term variability originates. A Keplerian orbit of 35~d corresponds to distances of $\sim${0.21}~AU to $\sim${0.33}~AU around stars of masses from $1\,M_\odot$ to $4\,M_\odot$, which indicates that longer-term variations on the star may be triggered {close to} the planet-forming regions of the disks \citep[e.g.,][]{dullemond2010,winn2015}. {Albeit at wider orbital distances}, hydrodynamical simulations of embedded, Jupiter-mass protoplanets orbiting solar-mass YSOs at distances $\simeq 10$~AU \citep{montesinos2015} have indeed shown that the radiative feedback from the planet can trigger order-of-magnitude changes in the local disk temperature and accretion rate.

\section{Summary and Conclusions} \label{sec:conclusions}

In this study, we have combined high-precision time series photometry from {\it K2} Campaign~9 with simultaneous multi-band monitoring, obtained from the ground with VST/OmegaCAM, to investigate the origin and characteristic timescales of variability exhibited by $\sim$1--2~Myr-old PMS stars in the Lagoon Nebula region. This investigation has allowed us to extend the existing framework for YSO variability studies, established in the course of previous analyses targeting the K to M spectral range, to early-type (B, A) stars, thereby revealing significant trends with stellar mass. We employed the {\it K2} light curves to identify distinct variability classes across our sample, based on the morphology, periodicity, and symmetry of the dominant flux variations observed case by case along the times series. We then mapped the color distribution of the different groups of variables on various photometric diagrams from the UV to the IR, in order to pinpoint the underlying physical drivers.

In agreement with previous studies on star-forming regions such as NGC~2264, $\rho$~Ophiucus, and Upper Scorpius, we identified distinct variability signatures for disk-dominated and photosphere-dominated sources. Irregular light curve patterns, like bursting, stochastic, or dipping, are observed on stars with large near- and mid-IR excesses, indicative of substantial material in the inner disk environment. Conversely, regular light curve patterns, like periodic {or} quasi-periodic, are typically associated with stars that exhibit IR colors similar to the photospheric levels, which indicates more evolved inner disks around these sources. The photometric properties measured in UV and H$\alpha$ filters indicate that the most erratic behaviors (bursting, stochastic) are driven by intense and/or time-variable accretion, while more modest and steady accretion activity is registered on stars with quasi-periodic variability features, either spot modulation or fading events (flux dips) driven by extended accretion curtains or inner disk structures.

While, overall, the vast majority ($\sim$90\%) of YSOs in our sample exhibit clear variability signatures in their {\it K2} light curves, around one third of the massive stars (B to F spectral types) do not show any definite variability pattern. A correlated variability analysis between optical and UV wavelengths further indicates that the statistical amount of variability displayed by stars B to early G is indistinguishable from the level of luminosity fluctuations measured for field stars, whereas young stars of later spectral types exhibit significant variability above the field population level. This gradient in the amount of variability measured on early-type vs. late-type YSOs is also observed within each individual {\it K2} morphological class, including flat-line light curves, where the intensity of luminosity fluctuations recorded around the typical flux level increases systematically from B and A stars to G and K stars. We suggest that these distinct variability properties for low-mass and high-mass young stars reflect a mass-dependent stellar magnetic field structure: simpler, largely dipolar and axisymmetric magnetic fields for late-type stars, which induce localized features at the stellar surface, as opposed to more complex and higher-order magnetic fields for early-type stars, which translate to a chaotic (and therefore homogeneous) distribution of surface starspots, with little variability in the spot-to-photosphere ratio at different rotational phases.

Early-type stars in our sample encompass a significantly lower fraction of disk-bearing sources than late-type stars, with a corresponding dearth of disk-dominated variability behaviors. In particular, while erratic light curves like stochastic and bursting are found among more massive (A, F) stars, we do not detect any instance of dipping behaviors at spectral types earlier than G. In the assumption that flux dips are driven by inner disk structures or dust entrained in accretion columns co-rotating with the star (as indicated by the similarity in intrinsic timescales of variability extracted for dipper light curves and for quasi-periodic light curves dominated by starspot modulation), the lack of dipping behaviors among early-type stars can be understood in terms of their high effective temperatures, which push the dust sublimation radius further out in the disk, well beyond the star--disk co-rotating regions. In addition, the absence of a strong dipolar field component in more massive stars would also hamper the development of a stable pattern of star--disk interaction, which is a prerequisite for quasi-periodic accretion features.

Finally, we investigated the dominant timescales of intrinsic variability emerging from the {\it K2} light curves. In agreement with previous studies, we found the day-to-week timescales to be the leading source of intrinsic variability for all YSOs on baselines of at least the months-long timescales probed within a single {\it K2} campaign. However, our results also suggest that distinct variability signatures within this time frame may arise from distinct locations within the star--inner disk environment. Stochastic stars, in particular, exhibit somewhat longer timescales of variability than other disk-dominated sources, and they stand out with their typical mid-IR colors, at the reddest end of the color locus traced by Class~II objects. These features suggest that their erratic behavior may be driven by time-variable accretion efficiency on the outer magnetospheric regions, beyond the co-rotation radius. Aperiodic dipper stars also exhibit timescales of variability comparable to those found for stochastic stars, and their respective color--luminosity trends indicate that these two classes of variables may be triggered by the same physical mechanism (variable mass load onto the accretion column, with consequent rapid evolution of the magnetospheric accretion structures), coupled with different viewing geometries. Burster stars, instead, exhibit typical timescales of variability consistent with those measured for quasi-periodic variables (modulated and dippers), suggesting an origin of the observed variability in the innermost disk regions co-rotating with the star. The large UV and H$\alpha$ excesses measured here for bursting variables confirm earlier physical explanations of their photometric behavior in terms of intense accretion activity driven by instabilities at the inner disk--magnetosphere boundary.

This study demonstrates the effectiveness of coordinated high-precision and multi-wavelength observing campaigns to identify the physical mechanisms that trigger the manifold variability patterns recorded for young stars, in relation to different stellar and circumstellar properties. Many unknowns still remain on how the physics and evolutionary timescales of individual YSOs is influenced by the surrounding environment, in particular by the presence of stellar companions. To address this issue, we have undertaken a targeted multiplicity survey of young stars in the Lagoon Nebula, guided by the {\it K2} analysis of variability presented here. Results from that survey, which will explore the mass dependence of the early multiplicity fraction and its connection to the star--disk dynamics, will be presented in a subsequent paper.

\acknowledgments
{We wish to thank the anonymous reviewer for a prompt report that helped improve the manuscript's clarity.} We gratefully acknowledge Janet Drew for making her narrow-band H$\alpha$ filter available to us for the VST/OmegaCAM observations. L.V.’s research was supported by an appointment to the NASA Postdoctoral Program at the NASA Ames Research Center, administered by Universities Space Research Association under contract with NASA. {The work was also supported by the National Aeronautics and Space Administration (NASA) under grant number 80NSSC21K0633 issued through the NNH20ZDA001N Astrophysics Data Analysis Program (ADAP).} This study employed data products from observations made with ESO Telescopes at the La Silla Paranal Observatory under programme ID 177.D-3023, as part of the VST Photometric H$\alpha$ Survey of the Southern Galactic Plane and Bulge (VPHAS+, www.vphas.eu). This publication also makes use of data products from the UKIRT Infrared Deep Sky Survey, the Two Micron All Sky Survey, and the Wide-field Infrared Survey Explorer. The Two Micron All Sky Survey is a joint project of the University of Massachusetts and the Infrared Processing and Analysis Center/California Institute of Technology, funded by the National Aeronautics and Space Administration and the National Science Foundation. The Wide-field Infrared Survey Explorer is a joint project of the University of California, Los Angeles, and the Jet Propulsion Laboratory/California Institute of Technology, funded by the National Aeronautics and Space Administration.

%

\vspace{5mm}
\facilities{Kepler/K2, VST (OmegaCAM), Spitzer (IRAC)}


\software{TOPCAT \citep{taylor2005}, NumPy \citep{oliphant2006}, Scikit-learn \citep{pedregosa2011}}





\bibliography{references}{}
\bibliographystyle{aasjournal}



\end{document}